\documentclass[sigconf]{acmart}

\usepackage{url}    
\usepackage{latexsym}
\usepackage{tablefootnote}
\usepackage{amsmath}
\usepackage{multirow}
\usepackage{enumerate}
\usepackage{graphicx}
\usepackage{subfigure}
\usepackage{graphicx}
\usepackage{threeparttable}
\usepackage{enumitem}
\usepackage{fancyhdr,graphicx}
\usepackage[ruled,vlined]{algorithm2e}
\newcommand{\eat}[1]{}

\AtBeginDocument{%
  \providecommand\BibTeX{{%
    \normalfont B\kern-0.5em{\scshape i\kern-0.25em b}\kern-0.8em\TeX}}}

\copyrightyear{2022}
\acmYear{2022}
\setcopyright{acmcopyright}

\acmConference[SIGIR '22] {Proceedings of the 45th International ACM SIGIR Conference on Research and Development in Information Retrieval}{July 11--15, 2022}{Madrid, Spain.}
\acmBooktitle{Proceedings of the 45th International ACM SIGIR Conference on Research and Development in Information Retrieval (SIGIR '22), July 11--15, 2022, Madrid, Spain}
\acmPrice{15.00}
\acmISBN{978-1-4503-8732-3/22/07}
\acmDOI{10.1145/3477495.3532031}

\SetKwInOut{Input}{input}
\SetKwInOut{Output}{output}

 \usepackage{cleveref} 
 \crefname{section}{Section}{Sections}
 \crefname{theorem}{Theorem}{Theorems}
 \crefname{lemma}{Lemma}{Lemmas}
 \crefname{equation}{Equation}{Equations}
 \crefname{proposition}{Proposition}{Propositions}
 \crefname{claim}{Claim}{Claims}
\crefname{appendix}{Appendix}{Appendices}
   \crefname{algorithm}{Algorithm}{Algorithms}
 \crefname{figure}{Figure}{Figures}
 \crefname{table}{Table}{Tables}
 \crefname{remark}{Remark}{Remarks}
 \crefname{definition}{Definition}{Definitions}
 \crefname{equatinon}{Equation}{Equations}
 \crefname{corollary}{Corollary}{Corollaries}

\allowdisplaybreaks

\usepackage{thmtools}
\usepackage{thm-restate}

\let \oldtextcircled \textcircled
\renewcommand{\textcircled}[1]{\oldtextcircled{\footnotesize #1}}

\usepackage{enumitem}
\setlist[itemize]{leftmargin=9mm}

\def \x{\mathbf{x}}

\def \z{\mathbf{z}}
\def \h{\mathbf{h}}

\def \f{\mathbf{f}}

\def \g{\mathbf{g}}

\def \BK{\mathbf{K}}

\def \BC{\mathbf{C}}

\def \BH{\mathbf{H}}

\def \BM{\mathbf{M}}

\def \BQ{\mathbf{Q}}

\def \BS{\mathbf{S}}

\def \BV{\mathbf{V}}
\def \BW{\mathbf{W}}
\def \BZ{\mathbf{Z}}

\usepackage{mathtools}

\begin{document}
\fancyhead{}

\title{Neighbour Interaction based Click-Through Rate Prediction via Graph-masked Transformer}

\author{Erxue Min}
\authornote{Work done during Erxue's internship at Tencent AI Lab.}
\email{erxue.min@gmail.com}
\affiliation{%
\institution{National Centre for Text Mining, Department of Computer Science, The University of Manchester}
\country{United Kingdom}
}

\author{Yu Rong}
\authornote{Corresponding author.}
\email{yu.rong@hotmail.com}
\author{Tingyang Xu}
\email{tingyangxu@tencent.com}
\author{Yatao Bian}
\email{yatao.bian@gmail.com}
\affiliation{%
\institution{Tencent AI Lab}
\country{China}
}

\author{Da Luo}
\email{lodaluo@tencent.com}
\author{Kangyi Lin}
\email{plancklin@tencent.com}
\affiliation{%
\institution{Weixin Open Platform, Tencent}
\country{China}
}

\author{Junzhou Huang}
\email{jzhuang@uta.edu}
\affiliation{%
\institution{Department of Computer Science and Engineering, University of Texas at Arlington}
\country{United States}
}

\author{Sophia Ananiadou}
\email{Sophia.Ananiadou@manchester.ac.uk}
\affiliation{%
\institution{National Centre for Text Mining, Department of Computer Science, The University of Manchester}
\country{United Kingdom}
}

\author{Peilin Zhao}
\email{masonzhao@tencent.com}
\affiliation{%
\institution{Tencent AI Lab}
\country{China}
}

\renewcommand{\shortauthors}{Erxue Min and Yu Rong, et al.}

\begin{abstract}
Click-Through Rate (CTR) prediction, which aims to estimate the probability that a user will click an item, is an essential component of online advertising. 
Existing methods mainly attempt to mine user interests from users' historical behaviours, which contain users' directly interacted items. Although these methods have made great progress, they are often limited by the recommender system's direct exposure and inactive interactions, and thus fail to mine all potential user interests. To tackle these problems,
we propose Neighbor-Interaction based CTR prediction (NI-CTR), which considers this task under a Heterogeneous Information Network (HIN) setting. In short,  Neighbor-Interaction based CTR prediction involves the local neighborhood of the target user-item pair in the HIN to predict their linkage. 
In order to guide the representation learning of the local neighbourhood, we further consider different kinds of interactions among the local neighborhood nodes from both explicit and implicit perspective, and propose a novel Graph-Masked Transformer (GMT) to effectively incorporates these kinds of interactions to produce highly representative embeddings for the target user-item pair.
Moreover, in order to improve model robustness against neighbour sampling, we enforce a consistency regularization loss over the neighbourhood embedding.
 We conduct extensive experiments on two real-world datasets with millions of instances and the experimental results show that our proposed method outperforms state-of-the-art CTR models significantly. Meanwhile, the comprehensive ablation studies verify the effectiveness of every component of our model. Furthermore, we have deployed this framework on the WeChat Official Account Platform with billions of users. The online A/B tests demonstrate an average CTR improvement of 21.9\%  against all online baselines.

\end{abstract}




\maketitle

\section{Introduction}
\label{intro}
In many applications such as online advertising and product search, Click-Through Rate (CTR) is a key indicator in business valuation, which measures the probability of a user clicking or interacting with a candidate item. For applications with a large user base, even a small improvement on CTR can potentially contribute to a large increase in the overall revenue.  However, achieving accurate CTR prediction remains a great challenge. This is due to the fact that the data in CTR prediction problems are usually of large scale and high sparsity, involving many categorical features of different fields. 

Typically, the data of CTR prediction is represented as high-dimensional and sparse categorical feature groups. To discover the potential click-through relation between user and item, the most popular learning paradigm is to firstly use an embedding layer to transfer the sparse user/item features to a low-dimensional dense embedding, and then construct the feature fusion \& learning models to encode the user preferences, item characteristic or their interactions. Typical models include Wide\&Deep \cite{cheng2016wide}, DeepFM \cite{guo2017deepfm}, xDeepFM \cite{lian2018xdeepfm}, AFM \cite{xiao2017attentional}, DeepMCP \cite{ouyang2019representation} and so on. However, this learning paradigm treats the sparse categorical feature equally and ignores the intrinsic structures among them, e.g., the sequential order of historical behaviors.

Recently, several studies in user interests modeling \cite{zhou2018deep,zhou2019deep,feng2019deep,lv2019sdm,lyu2020deep} emphasize on the sequential structure of user behaviour features. They model the historical items of users as sequences and exploit the sequence modeling methods such as LSTM \cite{hochreiter1997long}, GRU \cite{cho2014learning} and multi-head attention \cite{vaswani2017attention} to effectively model the user preference. Typical methods include DIN \cite{zhou2018deep}, DIEN \cite{zhou2019deep}, DSIN \cite{feng2019deep}, SDM \cite{lv2019sdm} and DMR \cite{lyu2020deep}, etc. 
Although existing methods for CTR prediction have achieved significant progress, the above methods only focus on mining the interaction between the candidate item and the user's historical behaviours, which suffers from two limitations:
On the one hand, user behaviours might be sparse for inactive users, which rise a cold-start problem and impede the quality of representation. On the other hand, due to the limitation of recommender system's exposure, the direct associated items of a user are not exhaustive enough to reflect all his/her potential interests.

To tackle these limitations, we propose Neighbour-Interaction based CTR (NI-CTR) prediction , which extends the prediction of a candidate user-item pair to their local neighbourhoods in a pre-defined Heterogeneous Information Network (HIN).
Specifically, we construct the HIN based on entities and relations which are associated with users and target items in the CTR prediction. Take video recommendation on WeChat Official Account as an example. In this task, we are interested in the CTR of videos. However, many additional information, such as users' click history of articles or news, the subscribe relation between users and Official Accounts, and the publish relation between official accounts and contents (videos/articles) can reflect the users' clicking preference and contribute important clues for inference. 
To make use of such rich side information, we construct the HIN based on these entities and relations. Afterwards, we leverage graph sampling methods to retrieve neighbours of both the target user and item in the HIN and integrate them to construct a merged local neighbourhood. 
In order to effectively learn the user-item clicking preference from the local neighborhood nodes,  we consider a wider range of interaction types between those nodes from the explicit and implicit perspectives,
 and propose to construct the four kinds of interaction graphs: 1) Induced graph edges in the HIN for modeling natural interaction, 2) Similarity graph for modeling node feature similarities, 3) Cross neighbourhood graph for capturing the interaction across user neighbours and item neighbours, and 4) Complete graph for modeling interaction between any nodes in the neighbourhood.

To better encode the information from those interaction graphs, 
we propose a novel Graph-masked Transformer (GMT) architecture to encode the neighbourhood, 
 which can flexibly involve structural
priors via a masking mechanism. 
In this way, the Transformer network is regularized by the graph priors and capable to learn more distinctive representations for the neighbourhood.
Besides, in order to reduce the noise introduced by sampling, we enforce a consistency regularization on the neighbourhood embedding, to make neighbourhood embeddings of an identical user-item pair similar.
Note that our neighbour interaction method is different from conventional graph-based methodologies such as HetGNN \cite{zhang2019heterogeneous} or HAN \cite{wang2019heterogeneous}, where graph neural network architectures such as GCN \cite{kipf2016semi} are exploited to compress features of neighbouring nodes in the graph into single embedding vector before making a prediction. In our case, both explicit and implicit interactions amongst neighbours of the user and item are fully captured, which mitigates the early summarization issue as introduced in \cite{qu2019end}. Moreover, the deep architecture of Transformer network enhances the feature interaction and improves the feature extraction capability. 
Furthermore, the heterogeneous graph setting enables the model to utilize cross-domain information, which can be particularly useful when direct interactions between users and items of the target type are scarce, somehow mitigating the Cold-start issue, which is verified in our experiments.
To sum up, our contributions to this paper can be concluded as follows:
\begin{itemize}[leftmargin=*]
\item We propose to exploit the neighbourhood of the target user-item pair in a HIN to assist the CTR prediction. Four types of interaction graphs are proposed to describe both explicit and implicit relations among the neighbours.
    \item We propose a novel Graph-Masked Transformer (GMT), which flexibly encodes
    topological priors into self-attention via a simple but effective graph masking mechanism.
    \item We propose a consistency regularization loss over the neighbourhood representation to alleviate the uncertainty of graph sampling, and thus improve the robustness of the model.
    \item We evaluate our method on both public and industrial datasets, demonstrating significant improvements of our methods over state-of-the-art methods in both CTR prediction and graph modeling. Furthermore, we have deployed our framework in the video service of WeChat Official Account Platform, and Online A/B tests also show that it outperforms existing online baselines by 21.9\%.
\end{itemize}

\begin{figure}
    \centering
    \includegraphics[width=0.8\linewidth]{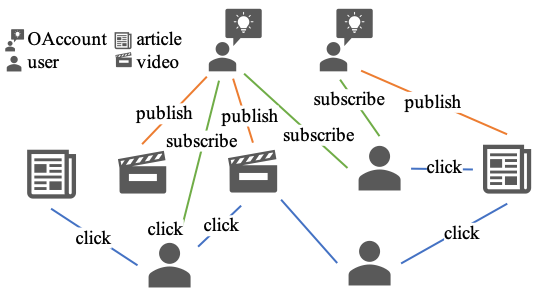}
    \caption{An illustration of the constructed HIN. It contains four kinds of nodes (OAccount, article, user and video) and three kinds of edges (click, publish and subscribe).}
    \label{figs:hin_example}
    \vspace{-3ex}
\end{figure}

\section{Preliminaries}

\subsection{CTR Data Modeling}
\label{ctr_feat}
Data in industrial CTR prediction tasks is mostly in a multi-group categorical form.
For example, [gender=Male,  visited\_categories=\{Sports\}, visited\_tag=\{Football, basketball\}], which are normally transformed into high-dimensional sparse binary features via encoding. Formally, given a user or an item , its feature vector can be represented as $[\f_1, \cdots, \f_k]$, where $\f_i \in \mathbb{R}^{d_{f_i}}$ is the encoding vector in the $i$-th feature group $\mathcal{F}_i$, $k$ is the total number of fields. Mathematically, we adopt the one/multi-hot representation with dimension $d_{f_i}$ to encode the categorical feature group, where $d_{f_i}$ is the number of unique ids in $\mathcal{F}_i$. $\f_i[j]$ is the $j$-th element of $\f_i$, $\sum^{d_{f_i}}_{j=1}\f_i[j]=p$, with $p=1$ refers to one-hot encoding and $p>1$ to multi-hot encoding.  For example,  
the aforementioned instance with three groups of features are illustrated as:
$$
\underbrace{[0,1]}_{\text {gender=Male}} \underbrace{[0, \ldots, 1, \ldots 0]}_{\text {visited\_categories=\{Sports\}}
} \underbrace{[0, \ldots, 1, \ldots, 1, \ldots 0]}_{\text{visited\_tag=\{Football, basketball\}}}.
$$
As the categorical features can be extremely high-dimensional and sparse for CTR prediction task, people usually apply an embedding layer which makes a linear transformation on each feature group to generate the low-dimensional dense representations, i.e., $\x_i=\BW_i\f_i, \BW_i \in \mathbb{R}^{d_{x_i} \times d_{f_i}}, d_{x_i}$ is much smaller than $d_{f_i}$. In this way, we can obtain the corresponding dense feature representation $[\x_1,\cdots, \x_k]$. 

The traditional CTR prediction models concern two types of entities: user and item. In this paper, we consider more than two kinds of entities and try to model the CTR problem under heterogeneous graph settings.
Assume the node type set is $\mathcal{T_V}$, the original feature vector of node $i$ which belongs to type $t(i) \in \mathcal{T_V}$ is denoted as: $\f_{t(i)}=[\f_{1}^{t(i)}, \cdots \f_{k_{t(i)}}^{t(i)}]$. The corresponding low-dimensional dense feature vector is denoted as $\x_{t(i)}=[\x_{1}^{t(i)}, \cdots, \x_{k_{t(i)}}^{t(i)}]$.\footnote{For the simplification of notation, the subscript of a vector can either be the node index or the node type.} It's worth to note that nodes of different types may share the same feature groups, e.g., different categories of item (e.g., video, article, product) might share the same tag scheme. Given two node types $t_a$ and $t_b$, we denote $g(t_a, t_b)$ as the indices of the shared feature group of $t_a$ and $t_b$ node types in $t_a$ node type.

\subsection{Heterogeneous Graph Construction}\label{sec.hin}
In this paper, we consider an undirected Heterogeneous Information Network $\mathcal{G}(\mathcal{N}, \mathcal{E},\mathcal{T}_\mathcal{V},\mathcal{T}_\mathcal{E})$, where $\mathcal{N}$ is the node set, $\mathcal{E} \in \mathcal{N} \times \mathcal{N}$ is the edge set,$\mathcal{T}_\mathcal{V}$ is the  node type set and $\mathcal{T}_\mathcal{E} \in \mathcal{T}_\mathcal{V} \times \mathcal{T}_\mathcal{V}$ is the edge type set.  $\mathcal{N}=\{\mathcal{U},\mathcal{I},\mathcal{S}_1,..,\mathcal{S}_{|\mathcal{T}_\mathcal{V}|-2}\}$, where $\mathcal{U}$ is the user set, $\mathcal{I}$ is the item set which we are interested in, $\mathcal{S}_i$ is $i$-th relevant entity set. We assume that the edge type is decided by start and end node for simplicity. Take the WeChat Video Recommendation Scenario as an example, we construct the HIN based on four types of nodes: user, video, article and official account (OAccount), and five types of edges: user-click-video, user-click-article, user-subscribe-OAccount, OAccount-publish-video and OAccount-publish-article, as illustrated in Figure \ref{figs:hin_example}. Our task can also be considered as to predict the linkage between a user node and a candidate video node.

\subsection{Problem Definition}
In this section, we formulate the CTR prediction task with necessary notations. There are a set of $M$ users $\mathcal{U}=\{u_1,u_2,\ldots,u_M\}$, a set of $N$ items $\mathcal{I}=\{v_1,v_2,\ldots,v_N\}$. The user-item interactions are denoted as a matrix $\mathcal{Y} \in \mathcal{R}^{M\times N}$, where $y_{uv}=1$ denotes user $u$ clicks item $v$ before, otherwise $y_{uv}=0$. Given the task-associated HIN  $\mathcal{G}(\mathcal{N}, \mathcal{E})$ as described in Section \ref{sec.hin}, we can sample a batch of neighbouring nodes $\mathcal{N}_{uv} \in \mathcal{N}$ for each candidate user $u$ and item $v$ pair. Each node $r \in \mathcal{N}$ is associated with a feature vector as described in Section \ref{ctr_feat}, so we denote the feature vector sets of the sampled $\mathcal{N}_{uv}$ as $\mathcal{F}_{uv}$. Besides, we denote context features (e.g., time, matching method, matching score) as $\mathbf{C}$. Therefore, one instance can be represented as:
\begin{equation}
\vspace{-0.5ex}
    \{\mathcal{F}_{uv}, \mathbf{C}\}.
\vspace{-0.5ex}
\end{equation}
The goal of NI-CTR prediction is to \emph{predict the probability that user $u$ will click item $v$ based on the neighbourhood and context features.}

\begin{figure*}
    \centering
    \includegraphics[width=0.7\linewidth]{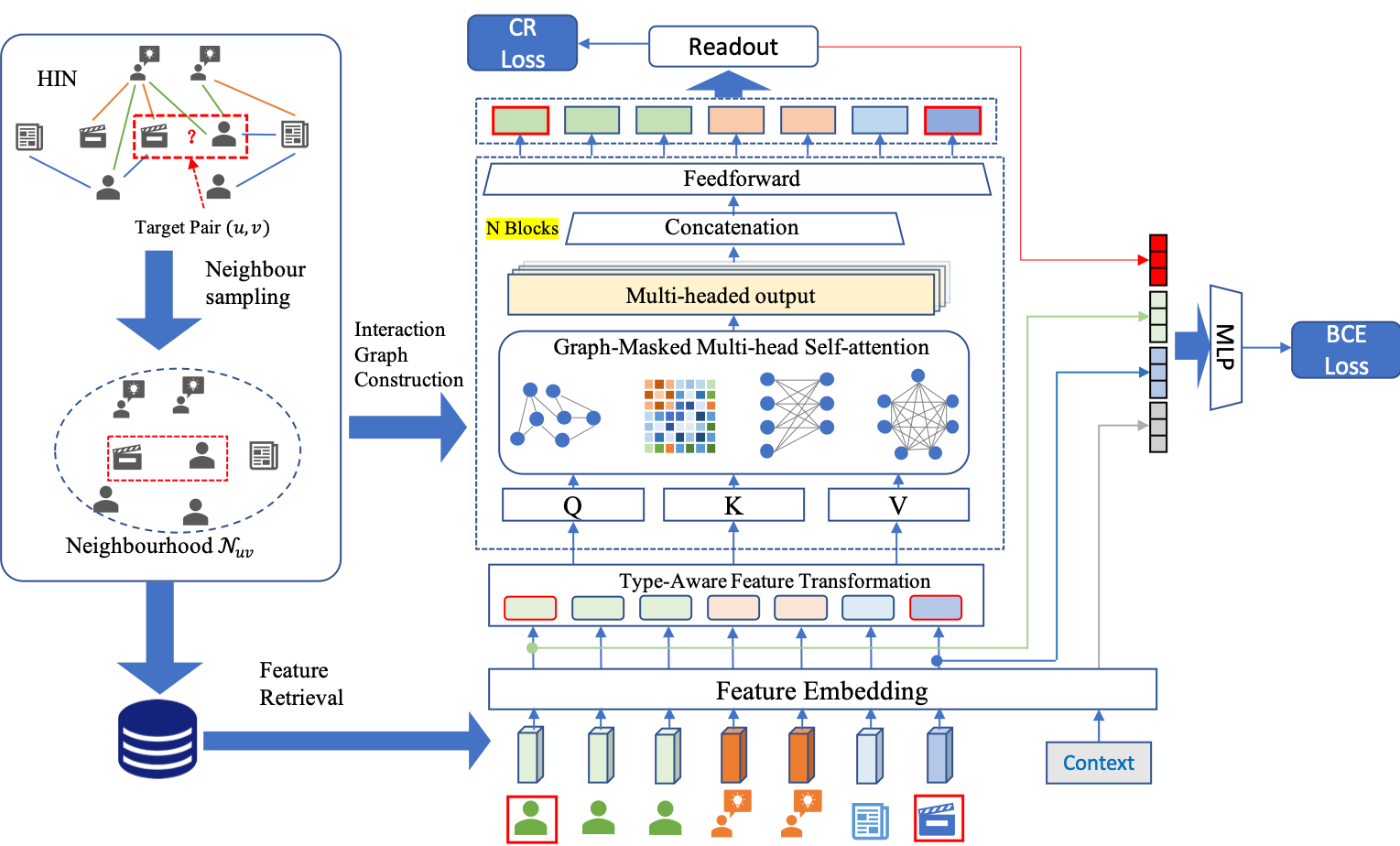}
    \vspace{-2ex}
    \caption{Overview of NI-CTR. Given a target user-item pair, we first perform neighbour sampling in the HIN to obtain associated neighbours. Then we retrieve the corresponding entity features and construct interaction graphs based on the neighbours. After that, we apply a Graph-Masked Transformer  to encode both the feature information and topological information. A binary cross-entropy loss and a consistency regularization loss are combined to optimize the network.}
    \vspace{-2ex}
    \label{fig:hmt_overview}
\end{figure*}

\section{Methodology}
In this section, we describe our framework to solve the NI-CTR prediction task. 
Figure~\ref{fig:hmt_overview}
 shows the overall framework: NI-CTR. It contains four main components: 1) Neighbour sampling in HIN, 2) Interaction graph construction on sampled nodes, 3) Graph-masked Transformer which encodes the neighbourhoods, 4) Loss and Optimization.  In the following sections, we will elaborate on the details of each component.

\subsection{Neighbour Sampling in HIN}
For each node $r$, there exist some relevant nodes in the pre-defined graph that may enrich its representation. Considering that the HIN sampling scenario in large-scale service and each node can be associated with rich features. We have the following requirements: 1) We should sample closest nodes as more as possible, because intuitively close nodes (e.g., one-hop neighbours) usually contains most relevant features/information, 2) We should sample nodes for each type in a pre-defined budget to limit the computational complexity, 3) We hope to sample nodes which have the most interactions (edges) with other nodes to maximize topological information.
To balance these requirements, we develop a simple but experimentally effective algorithm named Greedy Heterogeneous Neighbouring Sampling (GHNSampling). GHNSampling iteratively samples a list of nodes for a target node $r$ from one hop to further. Let $\{s_k\}_{k=1}^{|\mathcal{T_V}|}$ denotes the budget sampling sizes for each node type, $\mathcal{C}_k$ denotes the neighbours of type $k$ we have already sampled. GHNSampling greedily retrieves nodes from 1-hop to further until meeting the budget. In $l$-th hop, we retrieve  all the neighbours of nodes in $(l-1)$-th hop as $\mathcal{B}^l$, with $\mathcal{B}_k^l \subset \mathcal{B}^l$ as retrieved nodes of type $k$. 
For node $t \in \mathcal{B}^l$, we calculate the number of nodes it connects in the sampled node set $\mathcal{C}$ as $f_t=\vert \{s| (s,t) \in \mathcal{E}, s \in \mathcal{C}\}\vert$.
If $|\mathcal{B}_k^l|>s_k-|\mathcal{C}_k|$, we sample $s_k-|\mathcal{C}_k|$ nodes from $\mathcal{B}_k^l$ with the probability proportional to  $f_t$. We iteratively run the steps until budges of all node types are met.
This practice aims to retrieve nodes which have the most interactions with the target node to obtain topological information as much as possible. The details of GHNSampling are summarized in Algorithm \ref{sampling}.

\begin{algorithm}[t]
 \caption{GHNSampling}
 \label{sampling}
\SetKwInput{KwInput}{Input}                
\SetKwInput{KwOutput}{Output}              
\SetKwInput{KwInitial}{Initialization} 
\SetKwInput{KwResult}{Return}
\SetAlgoLined
\KwInput{HIN $\mathcal{G}(\mathcal{V}, \mathcal{E},\mathcal{T_V},\mathcal{T_E})$, sampling size for each type $\{s_k\}_{k=1}^{|\mathcal{T_V}|}$,
Target node $r$;
}
\KwOutput{Sampled node set $\mathcal{C}$} 
 \KwInitial{Initialize the node set $\mathcal{C}$ as $\{r\}$, $\mathcal{C}^0=\{r\}$}
 \For{$l$ in $1,2,3,\ldots$}{
 $\mathcal{B}^l=\{ \},\mathcal{C}^l=\{ \}$;\\
   retrieve all neighbours of nodes in $\mathcal{C}^{(l-1)}$ and add to buffer $\mathcal{B}^l$;\\
$\mathcal{B}^l = \mathcal{B}^l \setminus  \mathcal{C}$;\\
Count connection count $f_t$ for each distinct node $t$ in $\mathcal{B}^l$;\\
\For {node type $k \in |\mathcal{T_V}|$}
{Get $\mathcal{B}_k^l \subset \mathcal{B}^l,\mathcal{C}_k \subset \mathcal{C}$ ;\\
$n_k= \text{min}(|\mathcal{B}_k^l|,s_k-|\mathcal{C}_k|$);\\
randomly sample $n_k$ nodes from $\mathcal{B}_k^l$ with probability proportional to $f_t$ and add to $\mathcal{C}^l$;
}
$\mathcal{C}=\mathcal{C}\cup\mathcal{C}^l$;\\
\If{$|\mathcal{C}_k|=s_k$ \text{for any node type}  $k$}{Break
}}
\KwResult{$\mathcal{C}$}
\end{algorithm}
\vspace{-2ex}

\subsection{Construction of Local Interaction Graphs}
\label{inter_subgraphs}
\begin{figure}
    \centering
    \includegraphics[width=0.65\linewidth]{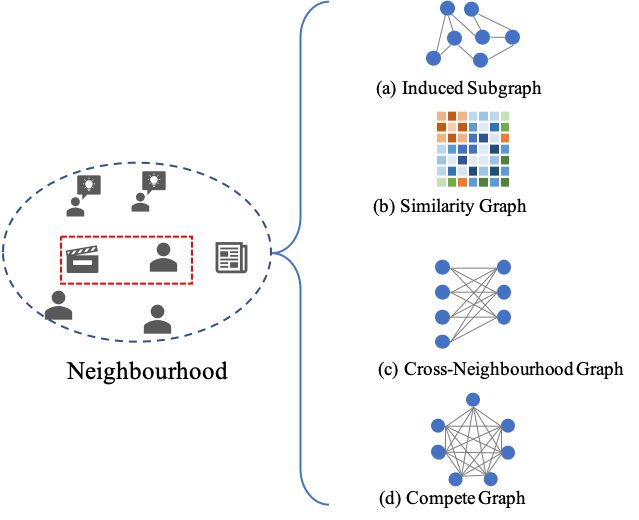}
    \vspace{-3ex}
    \caption{Four types of interaction graphs for neighbourhood modeling, which contain natural interactions, feature similarities, cross-neighbourhood interactions and all pair-wise interactions.}
    \vspace{-3ex}
    \label{fig:four_masks}
\end{figure}

After the neighbour sampling for the target user $u$ and candidate item $v$, we integrate their neighbours to obtain the neighbourhood of the $u$-$v$ pair, denoted as $\mathcal{N}_{uv}$ ($u,v \in \mathcal{N}_{uv}$). 
We associate each node $i$ in $\mathcal{N}_{uv}$ with its original feature vector $\f_i$. 
A direct solution to represent the sequence of nodes is to apply sophisticated models such as Transformer \cite{vaswani2017attention}, which consider the nodes in the neighbourhood as a complete graph, and learning representations based on the node features. However, to make the representation more distinctive and informative, in this section, we introduce four types of interaction graphs, as illustrated in  Figure~\ref{fig:four_masks}. The details of them are described as follows:

\textbf{Induced Subgraph} $\mathcal{G}^I_{uv}$: It is straightforward that the edge information in HIN provides important natural relation information among the nodes. Therefore, we retrieve all edges from HIN to generate the induced subgraph $\mathcal{G}^{I}_{uv}$.

\textbf{Similarity Subgraph $\mathcal{G}^{S}_{uv}$}: In the induced subgraph  $\mathcal{G}^{I}_{uv}$, only a subset of the categorical feature group which describes the behaviour relations or natural relations between different nodes are utilized to construct the graph. However, the other feature groups, such as item tags, which describes the rich latent semantic connections among nodes are ignored.
Although these node similarity relations can be implicitly captured by self-attention mechanism in Transformer, they will be decayed after the stacking of multiple layers, which might impede the performance.
Therefore, we define the similarity graph $\mathcal{G}^{S}_{uv}$ by the node feature similarities in the neighbourhood based on the original features of nodes.
We calculate all pairwise similarity scores as follows:
\begin{equation}
    \text{sim}(i,j) = \frac{\mathbf{f}_{i}[g(t(i),t(j))] \cdot \mathbf{f}_{j}[g(t(j),t(i))]}{\Vert \mathbf{f}_{i}[g(t(i),t(j))]\Vert \cdot \Vert\mathbf{f}_{j}[g(t(j),t(i))]\Vert},
\end{equation}
where $t(i)$ and $t(j)$ is the type of $i$- and $j$-th nodes. 
$\f_i$ is the original feature vector of node $i$, $g(t_a,t_b)$ is the feature group indices as mentioned in Section \ref{ctr_feat}.
Based on the similarity score, we investigate two approaches to construct the \emph{Similarity Graph} $\mathcal{G}_S$: (1) \textit{Weighted similarity graph}. we can directly construct the adjacent matrix $\BM_S$ based on similarity scores, i.e., $\BM_S[i,j]=\text{sim}(i,j)$. (2) \textit{$k$-NN similarity graph}. 
Although $\BS$ contains weights of similarities, it can be noisy due to data quality. Therefore, we apply a $k$-NN algorithm on it to retain only strong signals. Namely, $\BM_{\text{S}}[i,j]=1$ if $j$-th node is one of the $k$-nearest neighbours of $i$-th node.

\textbf{Cross Neighbourhood Subgraph $\mathcal{G}^{C}_{uv}$}: 
Although the $\mathcal{G}^{I}_{uv}$ and $\mathcal{G}^{S}_{uv}$ capture the natural relations and similarity relations of nodes in the neighbourhood. There are more implicit relations we should consider.
Let $\mathcal{N}_u$ and $\mathcal{N}_v$ denote the neighbours of $u$ and $v$ respectively, with $\mathcal{N}_{uv}=\mathcal{N}_u \cup \mathcal{N}_v$, we hope to capture all implicit interactions across the two neighbour set, which is inspired by Bidirectional attention flow for machine comprehension \cite{seo2016bidirectional}. We ensure $\mathcal{N}_{u} \cap \mathcal{N}_{v} = \emptyset$ by assigning overlapped nodes to the set with more connected edges.
Afterwards, we generate the cross-neighbour graph $\mathcal{G}^C_{uv}=\{(s,t)| s \in \mathcal{N}_u, t \in \mathcal{N}_v\}$.

\textbf{Complete Subgraph $\mathcal{G}^{P}_{uv}$}:
In this graph, we do not impose any structural prior and give the model the most freedom to learn any implicit correlations between nodes. The adjacent matrix of this graph is $\BM_P=\mathbf{1}_{|\mathcal{N}_{uv}| \times |\mathcal{N}_{uv}|}$.

\subsection{Graph-masked Transformer for Neighbourhood Interaction\&Representation}
After the construction of Local Interaction Graph, the neighbourhood of one instance can be represented as follows:
\begin{equation}
\vspace{-1ex}
    \{\mathcal{F}_{uv}, \mathcal{G}^I_{uv}, \mathcal{G}^S_{uv},\mathcal{G}^C_{uv},\mathcal{G}^P_{uv}\}.
\end{equation}
In order to learn a representation from both the node features and topological structure. We propose a novel Graph-masked Transformer (GMT), which basically consists of a Heterogeneous Node Feature Transformation layer, stacked Graph-masked Multi-head Self-attention layers and a readout layer. We will introduce the details of each module in the following section.

\subsubsection{Heterogeneous Node Feature Transformation layer}
For node $i$ in the neighbourhood $\mathcal{N}_{uv}$, we have its embedded dense feature vector $\x_i$ as described in Section \ref{ctr_feat}. Since nodes of different type have different feature groups and thus feature space, we use a type-aware feature transformation layer to embed them into a unified space:
\vspace{-1ex}
\begin{equation}
\h_i = \text{Linear}^{t(i)}(\x_i).
\end{equation}
where $t(i)$ is the node type of $i$-th node and $\text{Linear}^{t}(\cdot)$ is a linear layer of type $t$, with different types of layers have different trainable parameters.

\subsubsection{Graph-masked Multi-head Self-attention}
The key difference between GMT and the original Transformer architecture is in the Multi-head Self-Attention (MSA) layers. Given input sequence  $\BH=\{\h_1,\h_2,\ldots,\h_n\}$, the process of basic Self-attention mechanism can be defined as follows:
\vspace{-0.5ex}
\begin{align}
\vspace{-2ex}
    e_{ij} =\frac{(\BQ\h_i)^{\top}(\BK\h_i)}{\sqrt{d}},\\
     \alpha_{ij} = \frac{\exp(e_{ij})}{\sum_{k=1}^n\exp(e_{ik})},\\
         \z_i = \sum_{j=1}^n \alpha_{ij}(\BV\h_i),
\end{align}
where $\BQ,\BK,\BV$ are trainable parameter matrices, $d$ is the dimension of $h_i$, $n=\vert\mathcal{N}_{uv}\vert$ is the number of neighbours.
In a MSA layer, we have $H$ attention heads to implicitly attend to information from different representation subspaces of different nodes.
In our model, we attempt to use a \emph{graph-masking mechanism} to enforce the heads explicitly attend to different subspaces with graph priors. Specifically, we modify the calculation of the unnormalized attention score $e_{ij}$ as follows:
\begin{equation}
    e_{ij} =f_m(\frac{(\BQ\h_i)^{\top}(\BK\h_i)}{\sqrt{d}}, \BM_{ij}),
\end{equation}
where $\BM$ is the adjacent matrix of the prior graph, and $f_m(\cdot)$ is the masking function:
\vspace{-1ex}
\begin{equation}
f_m(x,\lambda)=
\begin{cases}
\lambda x& \lambda\neq 0\\
-\infty& \lambda=0.
\end{cases}
\end{equation}
This simple yet efficacious way enables the attention calculation aware of the structural priors.
Given the four types of interaction graphs described in Section \ref{inter_subgraphs}, we group heads into four sets and apply the graph masking with corresponding adjacent matrix. Now the output representation of $i$-th node $\h_i'$is computed as follows:
\begin{equation}
    \h_i' = \text{FFN}(\BW^O\text{Concat}(\z^{1}_{i}, \cdots,\z^{H}_{i})),
\end{equation}
where $\BW^O$ is the parameter matrix, $\text{FFN}(\cdot)$ is a two-layer feed forward layer with layer normalization \cite{ba2016layer} and residual connection \cite{he2016deep}.
With our Multi-head Graph-masked mechanism, we incorporate various graph priors into the Transformer architecture, which significantly extends the model expressivity.
After stacking multiple Graph-masked MSA layers, we have the final representation of nodes in the neighbourhood as: $\BZ = \{\z_1,\z_2,\ldots,\z_{|\mathcal{N}_{uv}|}\}$.

\subsubsection{Readout Layer}
To obtain a fixed-sized representation vector of the neighbourhood, we use a readout function:
\begin{equation}
    \g_{uv} = \text{Readout}(\BZ),
\end{equation}
in this paper, we simply use a mean pooling function to obtain the final neighbourhood embedding, i.e., $\g_{uv}=\frac{1}{\vert\mathcal{N}_{uv}\vert}\sum_{v_i \in \mathcal{N}_{uv}}\z_i$

\subsection{Classification and Optimization}
In this section, we introduce how to obtain the final prediction score and how to optimize the whole model. Although we have obtained neighbourhood embedding of the target $u$-$v$ pair, the feature information of the two targets: $u$ and $v$, despite being most valuable, might be somehow decayed after multiple layers of interaction with neighbours. To highlight the features of the two target nodes, we concatenate their initial dense embeddings $\x_u,\x_v$, with the neighbourhood embedding $\g_{uv}$ and the context features $\BC$. So the final embedding of the data instance is:
\vspace{-1ex}
\begin{equation}
    \z^o=\text{Concat}(\g_{uv},\x_u,\x_v,\BC).
\end{equation}
We adopt a MLP (Multi-Layer Perceptron) layer $f_{mlp}$ with parameter $\theta$ and a Sigmoid function $\sigma$ to predict the probability that the user $u$ will click the target item $v$:
\vspace{-1ex}
\begin{equation}
    \hat{y}_{uv} = \sigma(f_{mlp}(\z^o, \theta))
\end{equation}
Intuitively, in different training epochs, the neighbours sampled from the same $u$-$v$ pair can be different, which means the neighbourhood embedding $\g_{uv}$ and the final CTR $\hat{y}_{uv}$ would be different.
Assuming we sample maximum $S$ times for each $u$-$v$ pair, so the classification loss is the binary cross entropy loss on the training set $\mathcal{D}$:
\vspace{-1ex}
\begin{equation}
        \mathcal{L}_\text{BCE}=\frac{1}{ S}\sum_{<u,v> \in \mathcal{D}}\sum_{s=1}^{S}(y_{uv}\text{log}\hat{y}_{uv}^{s}+(1-y_{uv})\text{log}(1-\hat{y}_{uv}^{s})),
\end{equation}
where $\mathcal{D}$ is the training set, $y_{uv}\in\{0,1\}$ is the ground-truth label of the click-through.
In order to improve the robustness of models over sampling randomness, we propose a variant of consistency regularization \cite{feng2020graph}, which enforces the model to learn similar embeddings for the neighbourhoods sampled by the same $u$-$v$ pair. It can be formulated as:
\begin{equation}
\vspace{-1ex}
    \mathcal{L}_\text{CR} = \frac{1}{ S}\sum_{<u,v> \in \mathcal{D}}\sum_{s=1}^{S} \frac{1}{d_g}\vert\vert \hat{\g}_{uv}^s - \bar{\g}_{uv}^s \vert\vert
\end{equation},
where $\bar{\g}_{uv}^s =\frac{1}{S} \sum_{s=1}^{S}\hat{\g}_{uv}^s$, and $d_g$ is dimension of $\hat{\g}_{uv}^s$. The overall loss is obtained as a summation of both losses with coefficient $\gamma$:
\begin{equation}
\vspace{-1ex}
    \mathcal{L} = \mathcal{L}_\text{BCE}+\gamma\mathcal{L}_\text{CR}
    \vspace{-1ex}
\end{equation}.

\subsection{Complexity Analysis}
Since scalability is an important issue for CTR prediction in real-world industrial applications, we analyze the time complexity of GMT. The self-attention calculation is the main time cost, the computational complexity of which is $\mathcal{O}(n^2 \cdot d)$ where $d$ is hidden size of the neural network. In experiments, we found that we can achieve good performance with $n=100$. Another time cost is the calculation of similarity graph, which is $\mathcal{O}(n^2 \cdot d^{\hat{t}})$, where $d^{\hat{t}}$ is the average dimension of shared node features. Since the node features are usually very sparse, it can be efficiently calculated via sparse matrix libraries.

\section{Experiments}
We compare our method with baselines in both offline evaluation and online service. Results in both settings demonstrate the superiority of our methods.

\begin{table}[]
\caption{Statistics of the WeChat HIN}
\vspace{-1.5ex}
\resizebox{0.5\textwidth}{!}{
\begin{minipage}{0.48\textwidth}
\begin{tabular}{|l|l|l|l|}
\hline
Node type & Count & Fields & Features\footnote{Here we do not count in any entity (user/OAcc/article/video) ids, which would be extremely large.} \\ \hline
User      & 728M  & 75     & 147572   \\ \hline
OAcc  & 369K  & 95     & 187323   \\ \hline
Article   & 74M   & 26     & 148284   \\ \hline
Video     & 846K  & 23     & 134758   \\ \hline
\end{tabular}
\begin{tabular}{|l|l|l|l|}
\hline
Edge Type    & Count & Ave Src Deg & Ave Dst Deg \\ \hline
user-video   &  998M     &   1.35          &  1167.08           \\ \hline
user-article &  11.3B    &     15.53        &   151.25          \\ \hline
user-OAcc    &  33.9B     &   46.67          &   9726.76          \\ \hline
OAcc-video   &  50M     &    1.43         &   5.91          \\ \hline
OAcc-article &  74.7M     &   21.39          &   1.0          \\ \hline
\end{tabular}
\end{minipage}}
\label{node_edges}
\vspace{-3ex}
\end{table}

\subsection{Datasets}
Since there is no public large-scale CTR dataset that contains rich heterogeneous graph information, 
we build a new dataset from WeChat Video Recommendation System for CTR prediction. In our system, generally we have four types of nodes: User (U), Video (V), Article (A) and the content provider named Official Account (O), and five types of edges: user-click-video, user-click-article, user-subscribe-OAccount, OAccount-publish-Video and OAccount-publish-Article. We treat the graph as an undirected heterogeneous graph.
We use the log data from 14 consecutive days 
to construct the WeChat HIN, with over 0.8 billion nodes and 46 billion edges.
The detailed statistics are listed in Table \ref{node_edges}.
The dataset contains the log data of two successive days (Day\_1 and Day\_2)
with 20 million display/click logs of 17 million users and 0.5 million videos for each. Logs from Day\_1 are for training and logs from Day\_2 are for testing. This dataset is denoted as WC\_FULL. Besides, we also build a smaller dataset, using the logs of the first 12 hours of Day\_1 for training, while those of the left 12 hours for testing, with around 10 million logs for each. This dataset is denoted as WC\_SMALL.
 We would release the anonymized dataset for reproduction in the near future \footnote{A demo of data instances and the source codes are published in 
\textcolor{blue}{https://github.com/qwerfdsaplking/F2R-HMT}.}.
Meanwhile, we also evaluated our models on  Tmall\footnote{https://tianchi.aliyun.com/dataset/dataDetail?dataId=42}, which contains anonymized users' shopping logs in the past 6 months before and on the "Double 11" day. For each user, their clicked items are sorted by the interaction timestamp. We use the logs of a week before the "Double 11" day as the training set, and the logs on the "Double 11" day as the testing set. For each user, we randomly sample 10 non-clicked items to replace the target item as the negative samples.

\subsection{Competitors \& Metrics}
For offline evaluation, we compare our method with four categories of models: Feature Interaction (FI) models, User Interests Modeling (UIM) models, Graph Neural Networks (GNN) models, and Transformer-based models.
FI models include DeepFM \cite{guo2017deepfm} and xDeepFM \cite{lian2018xdeepfm}.
UIM models include DIN \cite{zhou2018deep}, DIEN \cite{zhou2019deep} and DMR \cite{lyu2020deep}. GNN models include GAT \cite{velivckovic2017graph}, GraphSAGE \cite{hamilton2017inductive}, RGCN \cite{schlichtkrull2018modeling}, HAN \cite{wang2019heterogeneous} and NIRec \cite{jin2020efficient}.
Transformer-based methods include Graph-Bert \cite{zhang2020graph} and Graph-Transformer \cite{dwivedi2020generalization}.
We adopt two widely-used evaluation metrics: $AUC$ and $Logloss$ \cite{guo2017deepfm}, to evaluate the offline performance. $AUC$ measures the goodness of assigning positive samples higher scores than randomly chosen negative samples. A higher $AUC$ value indicates a better performance. $Logloss$ measures the distance between the predicted scores and the ground-truth labels. A lower $Logloss$ value implies better performance.
For online evaluation, we use several online deployed CTR prediction models as baselines, including DeepFM \cite{guo2017deepfm}, GBDT\footnote{GBDT with DeepFM as feature extractor.} \cite{ke2017lightgbm}, DIN \cite{zhou2018deep}, DIEN \cite{zhou2019deep}, DeepMCP \cite{ouyang2019representation} and DMR \cite{lyu2020deep}.
We use the click rate to measure the performance of each method.

\subsection{Implementation Details}
For each target user-video instance, different categories of models have different forms of input features.
For FI models, we concatenate all features together as a vector. For UIM models, the inputs are the concatenated user-video features, along with a sequence of user behaviors.
In GNN models, the inputs are the two subgraphs of the target nodes, sampled based on their respective graph sampling models, and we use the final representation of the two target nodes for prediction. The inputs of Transformer-based models are the same as GMT, which are the neighbourhoods sampled by GHSampling. Since Graph-Transformer and Graph-Bert are designed for one type of edges, we choose the induced subgraph as their graph information.
For fair comparison, we set embedding dimension of all models as 100, the batch size as 128. We tune learning rate from \{1e-2,1e-3,1e-4,1e-5\}, dropout ratio from $0$ to $0.9$, hidden size of all deep layers from $\{64,100,128\}$, number of deep layers from $\{1,2,3,4\}$, maximum number of nodes in the subgraph from $\{50,100,200,400\}$, the regularization balancing coefficient $\gamma$ from $\{0.01,0.1,1\}$
We conduct a grid search for parameter selection.

\begin{table}[]
\caption{Results on offline datasets}
\vspace{-2ex}
\label{experiment1}
    \resizebox{1\linewidth}{!}{%
\begin{tabular}{l|l|ll|ll|ll}
\hline
\multirow{2}{*}{Category}    & \multirow{2}{*}{Model} & \multicolumn{2}{l|}{WC\_FULL}                                                            & \multicolumn{2}{l|}{WC\_SMALL}                                 & \multicolumn{2}{l}{Tmall}                                                                \\ \cline{3-8} 
                             &                        & \multicolumn{1}{l|}{AUC}                              & Logloss                          & \multicolumn{1}{l|}{AUC}    & Logloss                          & \multicolumn{1}{l|}{AUC}                              & Logloss                          \\ \hline
\multirow{2}{*}{FI}          & DeepFM                 & \multicolumn{1}{l|}{0.7009}                           & 0.2379                           & \multicolumn{1}{l|}{0.7022} & 0.2365                           & \multicolumn{1}{l|}{0.9012}                           & 0.1999                           \\
                             & xDeepFM                & \multicolumn{1}{l|}{0.7021}                           & 0.2370                           & \multicolumn{1}{l|}{0.7042} & 0.2354                           & \multicolumn{1}{l|}{0.9023}                           & 0.1978                           \\ \hline
\multirow{3}{*}{UIM}         & DIN                    & \multicolumn{1}{l|}{0.7042}                           & 0.2345                           & \multicolumn{1}{l|}{0.7073} & 0.2320                           & \multicolumn{1}{l|}{0.9034}                           & 0.1954                           \\
                             & DIEN                   & \multicolumn{1}{l|}{0.7043}                           & 0.2347                           & \multicolumn{1}{l|}{0.7069} & 0.2334                           & \multicolumn{1}{l|}{0.9045}                           & 0.1943                           \\
                             & DMR                    & \multicolumn{1}{l|}{0.7098}                           & 0.2280                           & \multicolumn{1}{l|}{0.7089} & 0.2310                           & \multicolumn{1}{l|}{0.9065}                           & 0.1926                           \\ \hline
\multirow{4}{*}{GNN}         & GraphSAGE              & \multicolumn{1}{l|}{0.7032}                           & 0.2366                           & \multicolumn{1}{l|}{0.7056} & 0.2378                           & \multicolumn{1}{l|}{0.9234}                           & 0.1789                           \\
                             & GAT                    & \multicolumn{1}{l|}{0.7130}                           & 0.2214                           & \multicolumn{1}{l|}{0.7145} & 0.2210                           & \multicolumn{1}{l|}{0.9245}                           & 0.1776                           \\
                             & RGCN                   & \multicolumn{1}{l|}{0.7078}                           & 0.2289                           & \multicolumn{1}{l|}{0.7101} & 0.2265                           & \multicolumn{1}{l|}{0.9201}                           & 0.1801                           \\
                             & HAN                    & \multicolumn{1}{l|}{0.7015}                           & 0.2378                           & \multicolumn{1}{l|}{0.7041} & 0.2399                           & \multicolumn{1}{l|}{0.9180}                           & 0.1823                           \\
                             & NIRec                    & \multicolumn{1}{l|}{0.7149}                           & 0.2200                           & \multicolumn{1}{l|}{0.7167} & 0.2197                           & \multicolumn{1}{l|}{0.9246}                           & 0.1775                           \\
\hline
\multirow{4}{*}{Transformer} & Transformer            & \multicolumn{1}{l|}{0.7200}                           & 0.2174                           & \multicolumn{1}{l|}{0.7260} & 0.2075                           & \multicolumn{1}{l|}{0.9339}                           & 0.1700                           \\
                             & Graph-Trans      & \multicolumn{1}{l|}{0.7201}                           & 0.2175                           & \multicolumn{1}{l|}{0.7277} & 0.2063                           & \multicolumn{1}{l|}{0.9321}                           & 0.1715                           \\
                             & Graph-BERT             & \multicolumn{1}{l|}{0.7211}                           & 0.2165                           & \multicolumn{1}{l|}{0.7290} & 0.2054                           & \multicolumn{1}{l|}{0.9345}                           & 0.1693                           \\
                             & GMT                    & \multicolumn{1}{l|}{\textbf{0.7290}} & \textbf{0.2103} & \multicolumn{1}{l|}{\textbf{0.7360}} & \textbf{0.2014} & \multicolumn{1}{l|}{\textbf{0.9410}} & \textbf{0.1603} \\ \hline
\end{tabular}}
\vspace{-2ex}
\end{table}

\subsection{Results on Offline Datasets}
Table \ref{experiment1} shows the experimental results on both WeChat dataset and Tmall. From Table \ref{experiment1}, we have the following observations:
\begin{itemize}[leftmargin=*]
    \item Our proposed GMT achieves obviously better performance and beat the other baselines on all datasets. It verifies that the interaction graphs of multiple types and the graph-masked mechanism empower the Transformer architecture to learn more informative representations.
    \item UIM methods are better than FI methods, demonstrating that user interest mining is useful for representation learning.
    \item GNN models, especially GAT and RGCN, are generally better than both FI and UIM methods. This is because they utilize the topological information of target nodes, which contains useful auxiliary information and contributes to a better feature interaction.
    \item The performance of HAN is obviously worse than other graph models. We conjecture that its shallow-layer architecture impairs its capability for node feature interaction.
    \item We notice that Transformer-based models with our data pipeline are significantly better than other categories of models. It can be attributed to the deep self-attention architecture, which has a more powerful representation capability and learns better feature interaction.
\end{itemize}

\subsection{Ablation Studies}
\subsubsection{Impacts of four types of interaction graph}
In this section, we investigate how each interaction graph in our model influences the final results. 
Table \ref{ablation} shows the results of removing or only keeping the masking matrix constructed by the specific interaction graph. We can find that:
\begin{itemize}[leftmargin=*]
    \item Only keeping a single type of masking matrix achieves obviously inferior performance than the full model (GMT), which means the models cannot learn sufficient structural information of input nodes.
    \item It is interesting that the model with only cross-subgraph masking performs better than that with fully-connected masking, which implies that the inter-subgraph information aggregation is more important than intra-subgraph information aggregation.
    \item Meanwhile, the removal of any masking matrix would downgrade the final performance. It demonstrates that each interaction graph contributes to the final results, improving the representation capability of the Transformer Network. 
\end{itemize}

\subsubsection{Impacts of consistency regularization}
In order to alleviate the noise and uncertainty of neighbour sampling, we enforce a consistency regularization on the embedding of neighbourhood generated by GMT. To verify its effectiveness, we remove this loss and compare it performance against our framework. As we can see in Table \ref{ablation} (Row 1 and Row 10), the performance of the model declines without the consistency regularization loss. Enforcing the neighbourhood embeddings of an identical $u$-$v$ pair to be similar, the model can be more robust and has a better performance.

\begin{table}[t]
\caption{Ablation results of each module on WC\_FULL dataset}
\vspace{-1.5ex}
\label{ablation}
\resizebox{0.36\textwidth}{!}{
\begin{tabular}{ccccl|cc}
\hline
\multicolumn{5}{c|}{Modules}                                                                                                                                                                                              & \multicolumn{2}{c}{WC\_FULL}                                      \\ \hline
$\mathcal{G}^{I}_{uv}$                                            & $\mathcal{G}^{S}_{uv}$                                            & $\mathcal{G}^{C}_{uv}$                                            & $\mathcal{G}^{P}_{uv}$                                            &       CR Loss                    & \multicolumn{1}{c|}{AUC}             & Logloss                    \\ \hline
\checkmark                     & \checkmark                     & \checkmark                     & \checkmark                     & \checkmark & \multicolumn{1}{c|}{\textbf{0.7290}} & \textbf{0.2103}            \\
\checkmark                     &                                               &                                               &                                               & \checkmark & \multicolumn{1}{c|}{0.7180}          & 0.2193                     \\
                                              & \checkmark                     &                                               &                                               & \checkmark & \multicolumn{1}{c|}{0.7179}          & 0.2193                     \\
                                              &                                               & \checkmark                     &                                               & \checkmark & \multicolumn{1}{c|}{0.7211}          & 0.2154                     \\
                                              &                                               &                                               & \checkmark                     & \checkmark & \multicolumn{1}{c|}{0.7203}          & 0.2157                     \\
                                              & \checkmark                     & \checkmark                     & \checkmark                     & \checkmark & \multicolumn{1}{c|}{0.7243}          & 0.2132                     \\
\checkmark                     &                                               & \checkmark                     & \checkmark                     & \checkmark & \multicolumn{1}{c|}{0.7252}          & 0.2126                     \\
\checkmark                     & \checkmark                     &                                               & \checkmark                     & \checkmark & \multicolumn{1}{c|}{0.7237}          & 0.2149                     \\
\checkmark                     & \checkmark                     & \checkmark                     &                                               & \checkmark & \multicolumn{1}{c|}{0.7263}          & 0.2123                     \\
\checkmark & \checkmark & \checkmark & \checkmark &                           & \multicolumn{1}{l|}{0.7274}          & 0.2116 \\
                          &                           &                           &                         &                           & \multicolumn{1}{l|}{0.7200}          & 0.2174 \\ \hline
\end{tabular}}
\vspace{-2ex}
\end{table}

\begin{figure}[t]
\centering
\includegraphics[width=0.7\columnwidth]{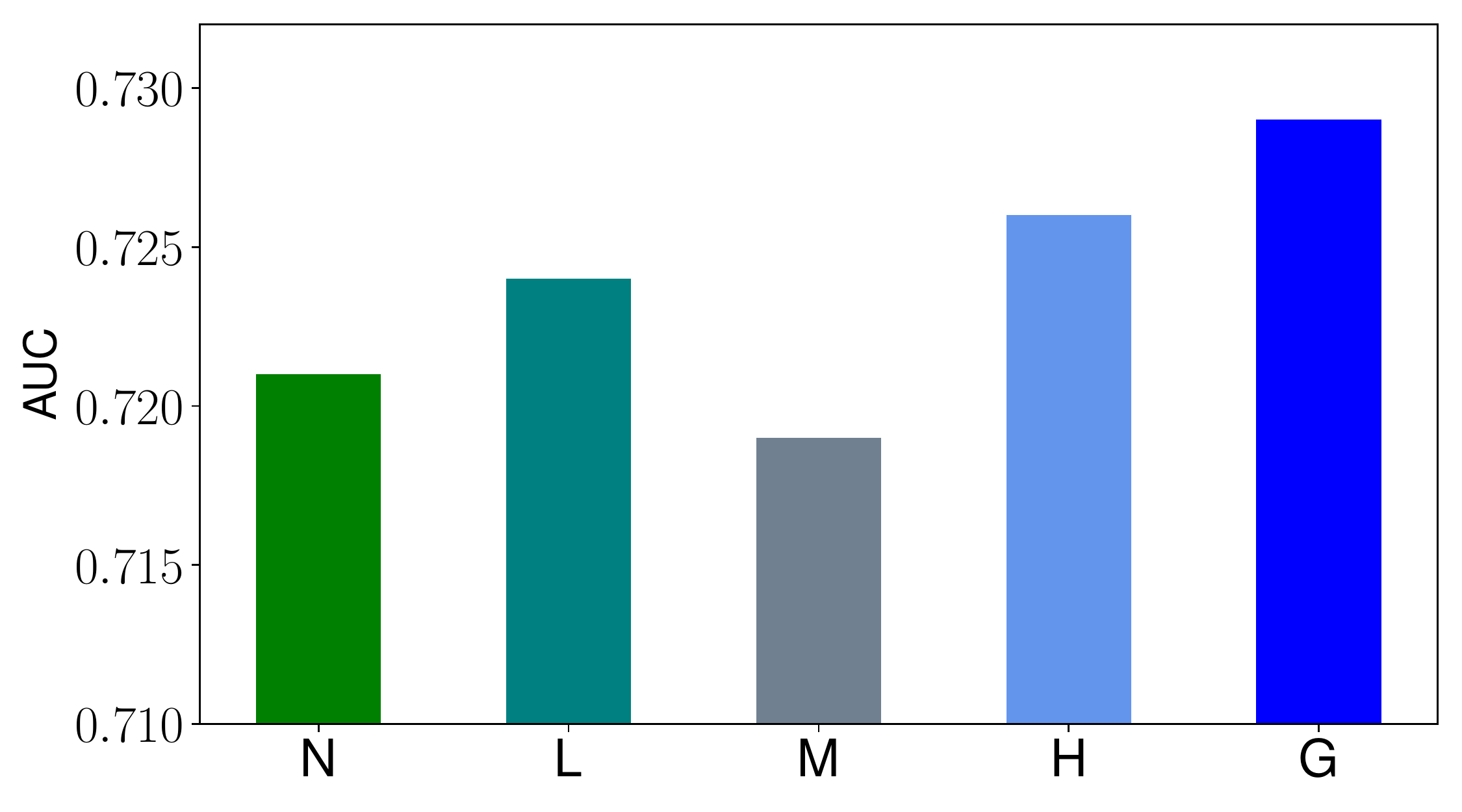}
\vspace{-3ex}
\caption{Results of graph sampling methods. \textbf{N}: Node-wise sampling; \textbf{L}: Layer-wise sampling; \textbf{M}: Metapath sampling; \textbf{H}: HGSampling; \textbf{G}: GHSampling.}
\vspace{-3ex}
\label{ctr_sampling}
\end{figure}

\begin{figure}[t]
\centering
\includegraphics[width=0.7\columnwidth]{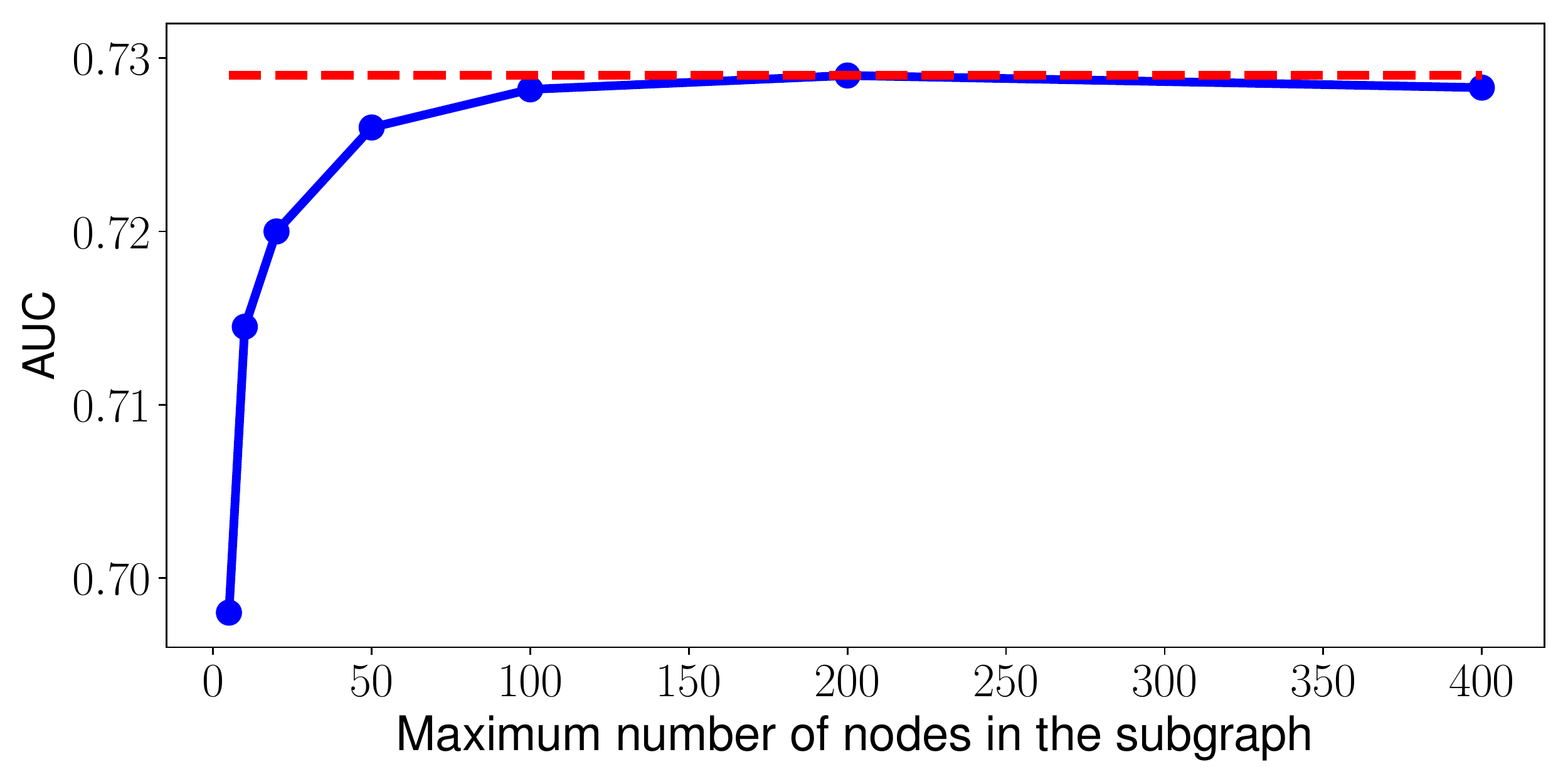}
\vspace{-3ex}
\caption{Results of different maximum numbers of sampled nodes in the subgraph.}
\vspace{-3ex}
\label{ctr_num}
\end{figure}

\begin{figure}[t]
\centering
\includegraphics[width=0.7\columnwidth]{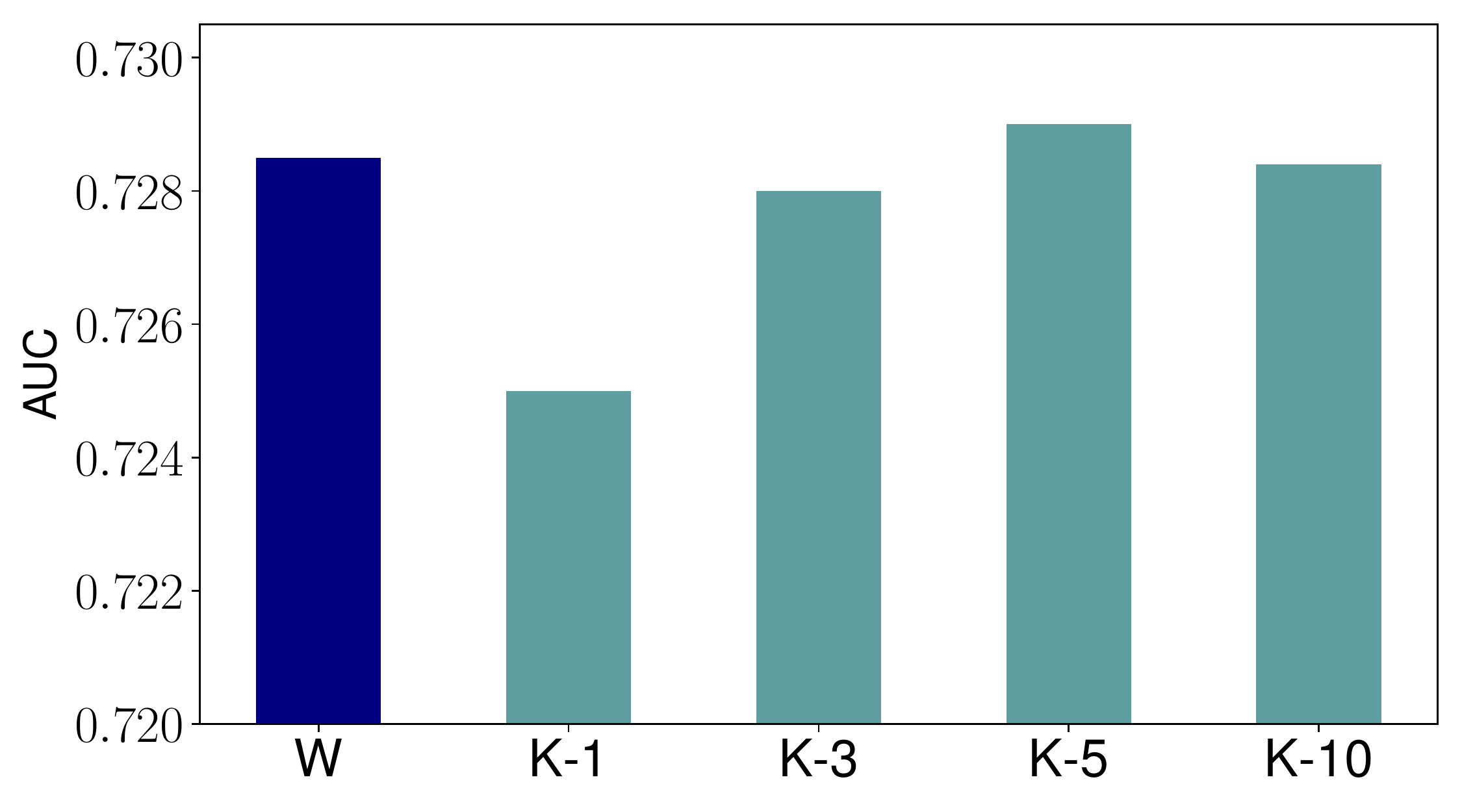}
\vspace{-3ex}
\caption{Results of different similarity graphs, where W denotes weighted similarity graph, and $K$-$n$ denotes $k$-NN similarity graph with $k=n$.}
\vspace{-3ex}
\label{ctr_sg1}
\end{figure}

\subsubsection{Impacts of subgraph sampling}
In this section, we compare the effects of subgraph sampling on WC\_FULL dataset. 
Firstly, we fix the sampling number $n_s$ of nodes in the subgraph as 200 and compare different graph sampling algorithms.
We implement four categories of graph sampling methods: 1) \textbf{N}ode-wise sampling \cite{hamilton2017inductive},with sample number of each node selected form $\{3,5,10\}$, 2) \textbf{L}ayer-wise sampling \cite{huang2018adaptive}, with layer number selected from $2,3,4$,  3) \textbf{M}etapath sampling \cite{jin2020efficient}, with three types of metapaths: UV,UOV,UAUV,UVUV, 4) \textbf{H}GSampling \cite{hu2020heterogeneous} with sample depth selected from $2,3,4$, 5) \textbf{G}HSampling. 

Figure \ref{ctr_sampling} shows the results of each sampling method with the best parameters. As we can see, layer-wise sampling and HGSampling are better than node-wise sampling and metapath sampling, and the greedy sampling strategy is superior to all methods. It is reasonable, as in such a HIN with rich and meaningful node features, the closest neighbours of a target node contain the most important and relevant information. Node-wise sampling and metapath sampling tend to sample further neighbors than close neighbours, which contains less useful information and introduce extra noise. On the contrary, the GHSampling method samples close neighbours as many as possible and maximizes the inner correlation, thus achieves the best performance.
Then we analyze the effects of $n_s$. As illustrated in Figure \ref{ctr_num}, the performance improves significantly when $n_s$ increases from $5$ to $100$, and reaches the top when $n_s$=$200$. Too large $n_s$ would consume heavy computational costs and impede the results.

\subsubsection{Impacts of similarity graph construction}
In this section, we compare the effects of the similarity Graph. We compare performance of weighted similarity graph with $k$-NN graph with $k$ selected from $\{1,3,5,10\}$. As illustrated in Figure \ref{ctr_sg1}, the performance of the weighted similarity graph is close to that of $k$-NN similarity graphs with fine-tuned $k$, and too small $k$ would impair the performance.

\begin{figure}[t]
\centering
\includegraphics[width=0.7\columnwidth]{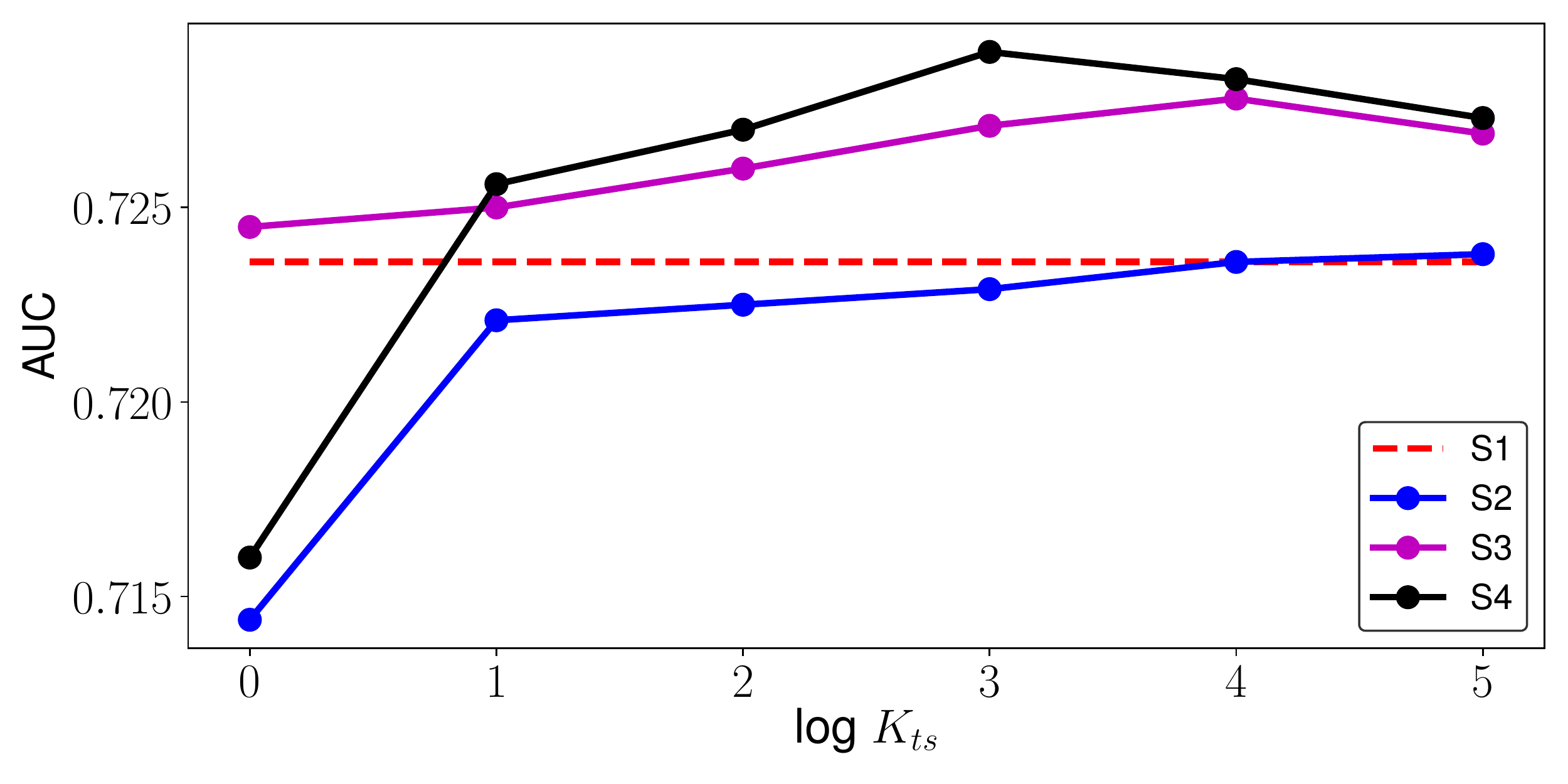}
\vspace{-3ex}
\caption{Results of different feature exploitation strategies with varied threshold value $K_{ts}$.}
\vspace{-3.5ex}
\label{ctr_sg2}
\end{figure}

\textbf{Feature exploitation strategies.} Moreover, we also conduct an in-depth analysis on four feature exploitation strategies: \textbf{S1}) Retain all node features, and do not use the similarity graph masking, \textbf{S2}) Remove feature groups with dimension $K>K_{ts}$ from all node features and do not use the similarity graph masking, \textbf{S3}) Retain all node features, and use feature groups with dimension $K>K_{ts}$ from node features to calculate similarity masking,  \textbf{S4}) Remove feature groups with dimension $K>K_{ts}$ from all node features, and use the removed features to calculate the similarity graph, where $K_{ts}$ is a threshold value and
we vary $K_{ts}$ from $10^0$ to $10^5$ to illustrate the variation in performance.
As shown in Figure \ref{ctr_sg2}, the
performance variation of \textbf{S2} shows that when $K_{ts}<10^5$, the more feature groups we use as node features, the better performance we have. However, we can also observe that removing feature groups with a dimension larger than $10^5$ in $\textbf{S2}$ has slightly better performance than that of \textbf{S1}, the baseline which uses all feature groups. It is because extremely sparse features might only bring marginal benefits while consume  more model parameters and introduce noise. \textbf{S2} shows that using a similarity graph can consistently bring advantages, but calculating the graph based on dense features might introduce noise and impair performance. $\textbf{S4}$ generally outperforms $\textbf{S3}$, demonstrating that some sparse features are more suitable to build connections between nodes. Removing them from node inputs improves both effectiveness and efficiency, which is a very practical trick in real service.

\begin{figure}[t]
\centering
\includegraphics[width=0.8\columnwidth]{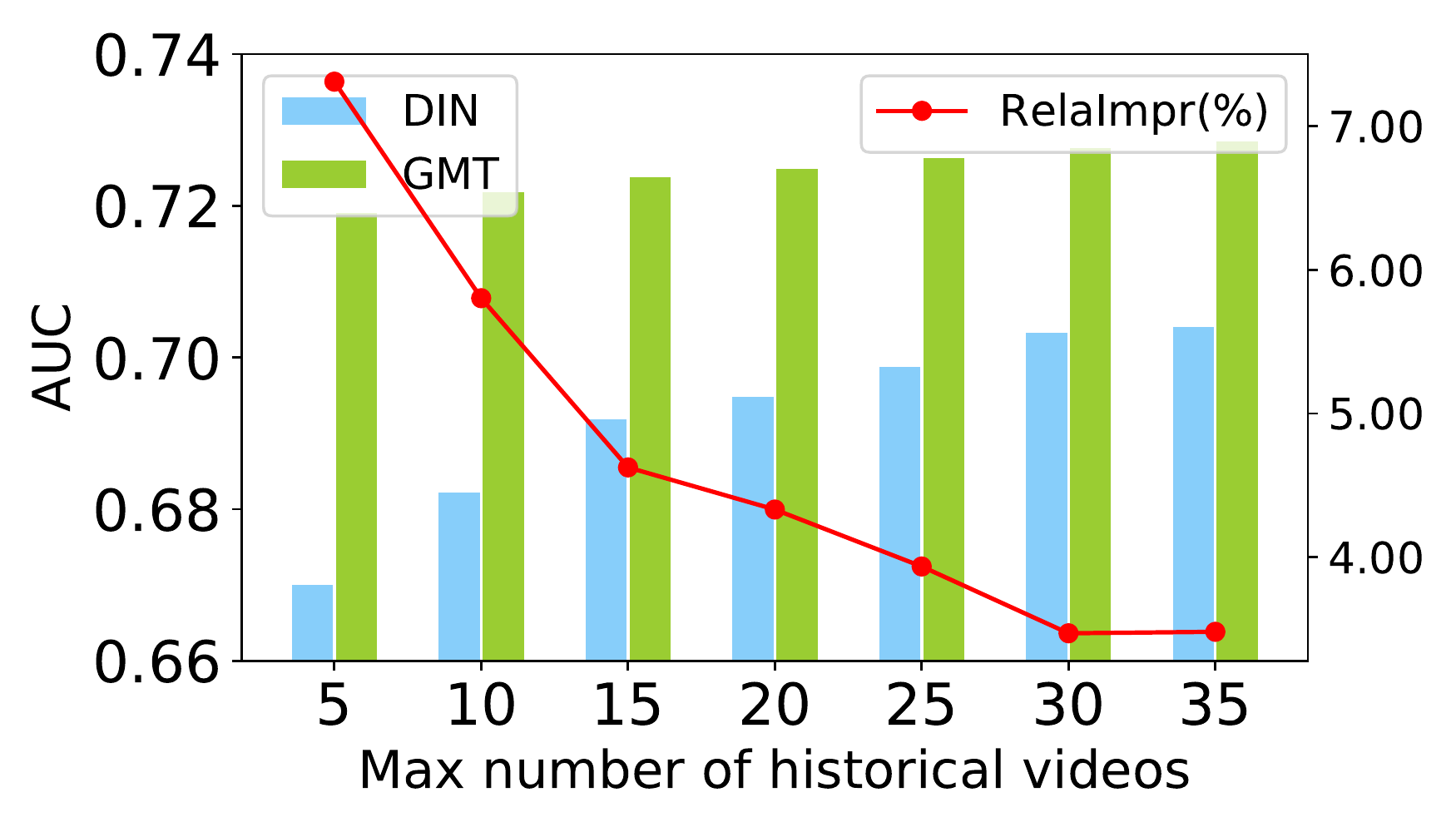}
\vspace{-3ex}
\caption{Cold-start analysis result.}
\vspace{-4ex}
\label{cold_start}
\end{figure}

\subsection{Performance in Cold-start Scenarios}
In this part, we conduct experiments on WC\_FULL dataset to verify that our model mitigates the Cold-Start problem.
Figure \ref{cold_start} shows the performance comparison between DIN and GMT in user cold-start scenarios with respect to different numbers of historical clicked videos for users.
The results in the figure illustrated that our method has a more significant relative improvement over DIN when the number of historical videos is smaller, which implies that GMT alleviates the Cold-start issue. The reason is that GMT makes use of extra heterogeneous graph information, which helps users to mine potential and implicit connections between users and videos.

\begin{figure}[t]
\centering
\includegraphics[width=0.85\columnwidth]{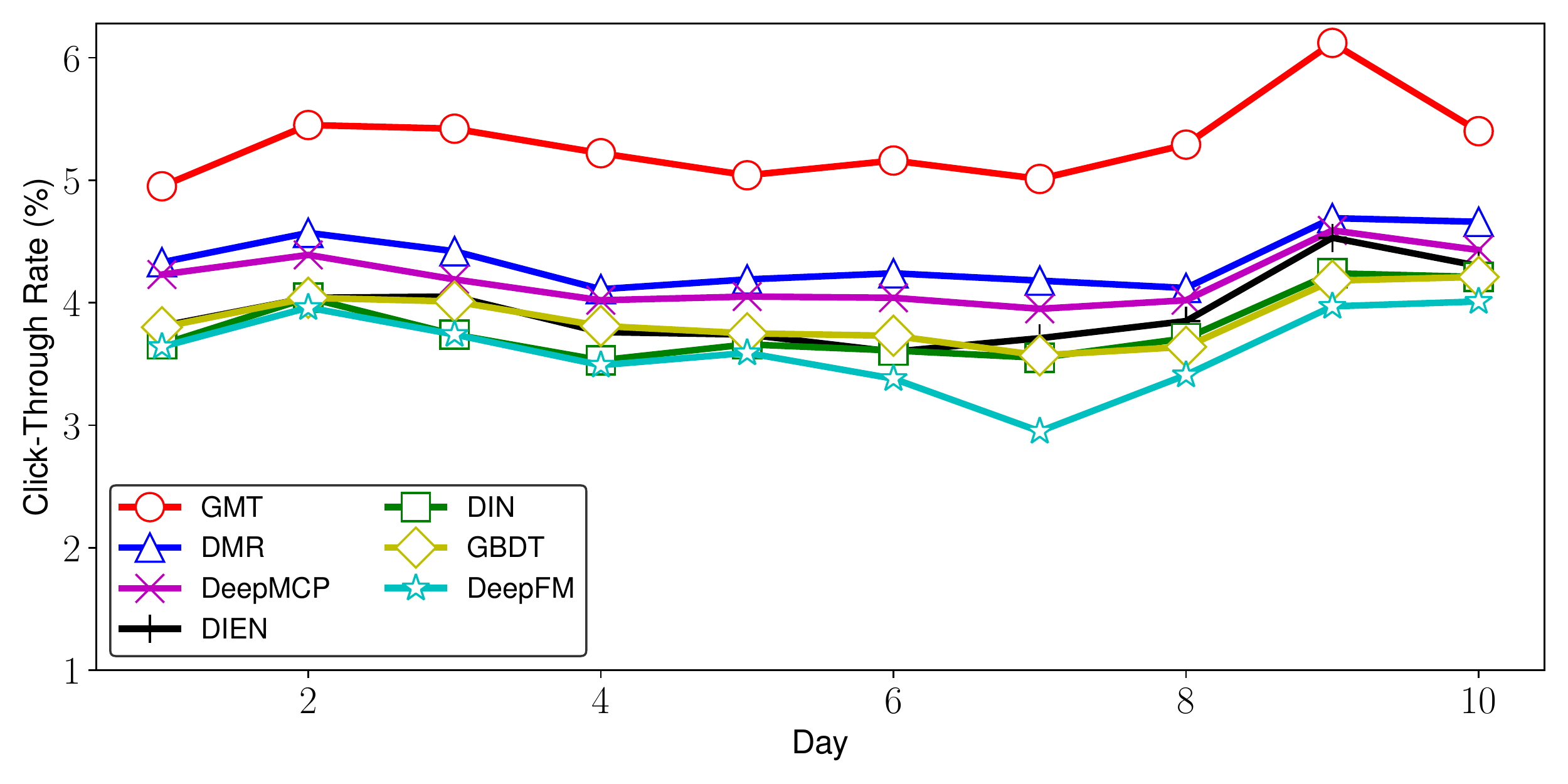}
\vspace{-3ex}
\caption{Results from Online A/B test during 10 consecutive days. The red curve is our method.}
\vspace{-3ex}
\label{online_results}
\end{figure}

\subsection{Online Serving \& A/B Testing}
We conduct online A/B Testing on the Video Recommendation Service of WeChat Official Account Platform during 10 consecutive days. 
Figure \ref{online_results} illustrates the results of our method with several baseline models. The results are collected from a consecutive 10 days. As we can see in the Figure, our proposed GMT outperforms other baselines significantly and it improves the best baseline by 21.9\% on average.
Note that other Top-5 models in the system mostly rely on extra techniques, such as multi-tasking training, pre-hash for item IDs, or node embedding pre-training, which further demonstrates the superiority and robustness of our methods.

\textbf{Deployment details} 
We save all the node relations into a graph database and update the database daily based on the behavior records of the latest day. We also update the features of each type of node daily based on the profiles of the latest day. We generate training datasets hourly based on the click behaviors in the latest hour. The model is continuously trained on the data of the latest 24 hours with 16 V100 GPUs. The number of training samples for 24 hours is around 40 million. One pass of all samples costs about one hour.

\section{Related Work}
\subsection{CTR Prediction}
CTR prediction has been extensively studied for many years. 
Due to the the high sparsity of input features, it is difficult to achieve good results via directly using raw features, and thus feature interaction modeling becomes a key role in this area \cite{rendle2010factorization,cheng2016wide,guo2017deepfm,wang2017deep,xiao2017attentional}. Factorization machines (FM) \cite{rendle2010factorization} use a low-dimensional vector to represent each feature field and learns 2-order feature interaction through inner product, achieving a significant improvement over linear models. 
Wide\&Deep \cite{cheng2016wide} combines a wide linear channel with cross-product and a deep neural network channel to capture feature interaction. DeepFM \cite{guo2017deepfm} integrates factorization machines and deep neural networks to learn the second-order crossover of features. xDeepFM \cite{lian2018xdeepfm} propose a novel compressed interaction network (CIN) to generate feature interactions in an explicit fashion and at the vector-wise level, combined with a classical DNN. DCN \cite{wang2017deep} adopts a multi-layer residual structure to learn higher-order feature representations. AFM \cite{xiao2017attentional} uses attention mechanism to automatically learn the weights of cross-features. 
Apart from learning embedding and interaction on handcrafted features, many works attempts to model user interests from user historical behaviors. Deep Interest Network (DIN) \cite{zhou2018deep} uses attention mechanism to assign different scores to user behaviors to learn the user representation. Deep Interest Evolution Network (DIEN) \cite{zhou2019deep} 
captures evolving user interest from their historical behaviors on items via a GRU network with attentional update gates. Deep Session Interest Network (DSIN) \cite{feng2019deep} 
leverages Bi-LSTM with self-attention layers to model the inter-session and intro-session interests of users.
However, although these models try to use powerful network architectures to model different kinds of historical behaviors, they did not make use of multi-source neighbourhood information, which limits their effectiveness.

\subsection{Graph Neural Networks for Recommendation}
Graph Neural Networks \cite{kipf2016semi,velivckovic2017graph,Rong2020DropEdge:,he2019cascade} have been widely explored in recommender systems in recent years, owing to their strong capability to model graph information in recommendation. GraphRec \cite{fan2019graph} makes the first attempt to introduce GNNs to social recommendation by modeling the user-item and user-user interactions as graph data. Wu $et al$. \cite{wu2019dual} propose a dual graph attention network to collaboratively learn representations for two-fold social effects. 
KGAT \cite{wang2019kgat} combines user-item graph with knowledge graph and uses graph convolution to obtain the final node representations. Heterogeneous graph Attention Network (HAN) \cite{wang2019heterogeneous} utilizes a semantic-level attention network and a node-level attention network to discriminate the importance of neighbor nodes and node types. HetGNN \cite{zhang2019heterogeneous} groups heterogeneous neighbours based on node types and use two modules to aggregate information from them.
Despite the progress, it is challenging to directly apply Graph-based recommendation methods to CTR prediction tasks for feature sparsity issues.

\subsection{Transformers for Graph Data}
There are some attempts to use Transformers in the context of graph-structured data. For example, GTransformer \cite{rong2020self} uses a tailed GNN layer as encoder to extract embeddings and feed them into Transformer. Graph Transformer \cite{dwivedi2020generalization} constrain the self-attention mechanism to local neighbourhoods of each node only.
Graph-BERT \cite{zhang2020graph} introduces three types of Positional Encoding to embed the node position information to the model.
Graphormer \cite{ying2021transformers} utilizes centrality encoding to enhance the node feature and uses spatial encoding along with edge encoding to incorporate structural inductive bias into the attention mechanism. Although these models have made great progress, they assume that the graphs are homogeneous and only have one type of edges, thus their performances are limited in our setting.


\section{Conclusion}
In this paper, we focus on exploiting the neighbourhood information to improve the performance of CTR prediction. We put CTR prediction into a heterogeneous graph setting and attempt to model the neighbourhood interaction. We propose four types of interaction graphs and design a novel model: Graph-masked Transformer to encode such interactions. Besides, we also design a consistency regularization to enhance the model robustness. Extensive experiments, including detailed ablation studies, verify the effectiveness of the neighbour interaction graph modeling and demonstrates the superiority and the flexibility of the proposed Graph-masked Transformer. We also successfully deploy the whole framework into an industrial scenario with billions of users and items and gain a significant improvement in CTR in a real application.  


\section{Acknowledgement}
Erxue Min acknowledges the support from The University of Manchester - China Scholarship Council joint scholarship. We thank the support from Weixin Open Platform and Weixin PlatoDB on the model online deployment.


\bibliographystyle{ACM-Reference-Format}
\balance
\bibliography{acmart}


\begin{thebibliography}{39}


\ifx \showCODEN    \undefined \def \showCODEN     #1{\unskip}     \fi
\ifx \showDOI      \undefined \def \showDOI       #1{#1}\fi
\ifx \showISBNx    \undefined \def \showISBNx     #1{\unskip}     \fi
\ifx \showISBNxiii \undefined \def \showISBNxiii  #1{\unskip}     \fi
\ifx \showISSN     \undefined \def \showISSN      #1{\unskip}     \fi
\ifx \showLCCN     \undefined \def \showLCCN      #1{\unskip}     \fi
\ifx \shownote     \undefined \def \shownote      #1{#1}          \fi
\ifx \showarticletitle \undefined \def \showarticletitle #1{#1}   \fi
\ifx \showURL      \undefined \def \showURL       {\relax}        \fi
\providecommand\bibfield[2]{#2}
\providecommand\bibinfo[2]{#2}
\providecommand\natexlab[1]{#1}
\providecommand\showeprint[2][]{arXiv:#2}

\bibitem[\protect\citeauthoryear{Ba, Kiros, and Hinton}{Ba
  et~al\mbox{.}}{2016}]%
        {ba2016layer}
\bibfield{author}{\bibinfo{person}{Jimmy~Lei Ba}, \bibinfo{person}{Jamie~Ryan
  Kiros}, {and} \bibinfo{person}{Geoffrey~E Hinton}.}
  \bibinfo{year}{2016}\natexlab{}.
\newblock \showarticletitle{Layer normalization}.
\newblock \bibinfo{journal}{\emph{arXiv preprint arXiv:1607.06450}}
  (\bibinfo{year}{2016}).
\newblock


\bibitem[\protect\citeauthoryear{Cheng, Koc, Harmsen, Shaked, Chandra, Aradhye,
  Anderson, Corrado, Chai, Ispir, et~al\mbox{.}}{Cheng et~al\mbox{.}}{2016}]%
        {cheng2016wide}
\bibfield{author}{\bibinfo{person}{Heng-Tze Cheng}, \bibinfo{person}{Levent
  Koc}, \bibinfo{person}{Jeremiah Harmsen}, \bibinfo{person}{Tal Shaked},
  \bibinfo{person}{Tushar Chandra}, \bibinfo{person}{Hrishi Aradhye},
  \bibinfo{person}{Glen Anderson}, \bibinfo{person}{Greg Corrado},
  \bibinfo{person}{Wei Chai}, \bibinfo{person}{Mustafa Ispir}, {et~al\mbox{.}}}
  \bibinfo{year}{2016}\natexlab{}.
\newblock \showarticletitle{Wide \& deep learning for recommender systems}. In
  \bibinfo{booktitle}{\emph{Proceedings of the 1st workshop on deep learning
  for recommender systems}}. \bibinfo{pages}{7--10}.
\newblock


\bibitem[\protect\citeauthoryear{Cho, Van~Merri{\"e}nboer, Gulcehre, Bahdanau,
  Bougares, Schwenk, and Bengio}{Cho et~al\mbox{.}}{2014}]%
        {cho2014learning}
\bibfield{author}{\bibinfo{person}{Kyunghyun Cho}, \bibinfo{person}{Bart
  Van~Merri{\"e}nboer}, \bibinfo{person}{Caglar Gulcehre},
  \bibinfo{person}{Dzmitry Bahdanau}, \bibinfo{person}{Fethi Bougares},
  \bibinfo{person}{Holger Schwenk}, {and} \bibinfo{person}{Yoshua Bengio}.}
  \bibinfo{year}{2014}\natexlab{}.
\newblock \showarticletitle{Learning phrase representations using RNN
  encoder-decoder for statistical machine translation}.
\newblock \bibinfo{journal}{\emph{arXiv preprint arXiv:1406.1078}}
  (\bibinfo{year}{2014}).
\newblock


\bibitem[\protect\citeauthoryear{Dwivedi and Bresson}{Dwivedi and
  Bresson}{2020}]%
        {dwivedi2020generalization}
\bibfield{author}{\bibinfo{person}{Vijay~Prakash Dwivedi} {and}
  \bibinfo{person}{Xavier Bresson}.} \bibinfo{year}{2020}\natexlab{}.
\newblock \showarticletitle{A Generalization of Transformer Networks to
  Graphs}.
\newblock \bibinfo{journal}{\emph{arXiv preprint arXiv:2012.09699}}
  (\bibinfo{year}{2020}).
\newblock


\bibitem[\protect\citeauthoryear{Fan, Ma, Li, He, Zhao, Tang, and Yin}{Fan
  et~al\mbox{.}}{2019}]%
        {fan2019graph}
\bibfield{author}{\bibinfo{person}{Wenqi Fan}, \bibinfo{person}{Yao Ma},
  \bibinfo{person}{Qing Li}, \bibinfo{person}{Yuan He}, \bibinfo{person}{Eric
  Zhao}, \bibinfo{person}{Jiliang Tang}, {and} \bibinfo{person}{Dawei Yin}.}
  \bibinfo{year}{2019}\natexlab{}.
\newblock \showarticletitle{Graph neural networks for social recommendation}.
  In \bibinfo{booktitle}{\emph{The World Wide Web Conference}}.
  \bibinfo{pages}{417--426}.
\newblock


\bibitem[\protect\citeauthoryear{Feng, Zhang, Dong, Han, Luan, Xu, Yang,
  Kharlamov, and Tang}{Feng et~al\mbox{.}}{2020}]%
        {feng2020graph}
\bibfield{author}{\bibinfo{person}{Wenzheng Feng}, \bibinfo{person}{Jie Zhang},
  \bibinfo{person}{Yuxiao Dong}, \bibinfo{person}{Yu Han},
  \bibinfo{person}{Huanbo Luan}, \bibinfo{person}{Qian Xu},
  \bibinfo{person}{Qiang Yang}, \bibinfo{person}{Evgeny Kharlamov}, {and}
  \bibinfo{person}{Jie Tang}.} \bibinfo{year}{2020}\natexlab{}.
\newblock \showarticletitle{Graph Random Neural Network for Semi-Supervised
  Learning on Graphs}.
\newblock \bibinfo{journal}{\emph{arXiv preprint arXiv:2005.11079}}
  (\bibinfo{year}{2020}).
\newblock


\bibitem[\protect\citeauthoryear{Feng, Lv, Shen, Wang, Sun, Zhu, and Yang}{Feng
  et~al\mbox{.}}{2019}]%
        {feng2019deep}
\bibfield{author}{\bibinfo{person}{Yufei Feng}, \bibinfo{person}{Fuyu Lv},
  \bibinfo{person}{Weichen Shen}, \bibinfo{person}{Menghan Wang},
  \bibinfo{person}{Fei Sun}, \bibinfo{person}{Yu Zhu}, {and}
  \bibinfo{person}{Keping Yang}.} \bibinfo{year}{2019}\natexlab{}.
\newblock \showarticletitle{Deep session interest network for click-through
  rate prediction}.
\newblock \bibinfo{journal}{\emph{arXiv preprint arXiv:1905.06482}}
  (\bibinfo{year}{2019}).
\newblock


\bibitem[\protect\citeauthoryear{Guo, Tang, Ye, Li, and He}{Guo
  et~al\mbox{.}}{2017}]%
        {guo2017deepfm}
\bibfield{author}{\bibinfo{person}{Huifeng Guo}, \bibinfo{person}{Ruiming
  Tang}, \bibinfo{person}{Yunming Ye}, \bibinfo{person}{Zhenguo Li}, {and}
  \bibinfo{person}{Xiuqiang He}.} \bibinfo{year}{2017}\natexlab{}.
\newblock \showarticletitle{DeepFM: a factorization-machine based neural
  network for CTR prediction}.
\newblock \bibinfo{journal}{\emph{arXiv preprint arXiv:1703.04247}}
  (\bibinfo{year}{2017}).
\newblock


\bibitem[\protect\citeauthoryear{Hamilton, Ying, and Leskovec}{Hamilton
  et~al\mbox{.}}{2017}]%
        {hamilton2017inductive}
\bibfield{author}{\bibinfo{person}{William~L Hamilton}, \bibinfo{person}{Rex
  Ying}, {and} \bibinfo{person}{Jure Leskovec}.}
  \bibinfo{year}{2017}\natexlab{}.
\newblock \showarticletitle{Inductive representation learning on large graphs}.
\newblock \bibinfo{journal}{\emph{arXiv preprint arXiv:1706.02216}}
  (\bibinfo{year}{2017}).
\newblock


\bibitem[\protect\citeauthoryear{He, Xie, Rong, Huang, Huang, Ren, and
  Shahabi}{He et~al\mbox{.}}{2019}]%
        {he2019cascade}
\bibfield{author}{\bibinfo{person}{Chaoyang He}, \bibinfo{person}{Tian Xie},
  \bibinfo{person}{Yu Rong}, \bibinfo{person}{Wenbing Huang},
  \bibinfo{person}{Junzhou Huang}, \bibinfo{person}{Xiang Ren}, {and}
  \bibinfo{person}{Cyrus Shahabi}.} \bibinfo{year}{2019}\natexlab{}.
\newblock \showarticletitle{Cascade-bgnn: Toward efficient self-supervised
  representation learning on large-scale bipartite graphs}.
\newblock \bibinfo{journal}{\emph{arXiv preprint arXiv:1906.11994}}
  (\bibinfo{year}{2019}).
\newblock


\bibitem[\protect\citeauthoryear{He, Zhang, Ren, and Sun}{He
  et~al\mbox{.}}{2016}]%
        {he2016deep}
\bibfield{author}{\bibinfo{person}{Kaiming He}, \bibinfo{person}{Xiangyu
  Zhang}, \bibinfo{person}{Shaoqing Ren}, {and} \bibinfo{person}{Jian Sun}.}
  \bibinfo{year}{2016}\natexlab{}.
\newblock \showarticletitle{Deep residual learning for image recognition}. In
  \bibinfo{booktitle}{\emph{Proceedings of the IEEE conference on computer
  vision and pattern recognition}}. \bibinfo{pages}{770--778}.
\newblock


\bibitem[\protect\citeauthoryear{Hochreiter and Schmidhuber}{Hochreiter and
  Schmidhuber}{1997}]%
        {hochreiter1997long}
\bibfield{author}{\bibinfo{person}{Sepp Hochreiter} {and}
  \bibinfo{person}{J{\"u}rgen Schmidhuber}.} \bibinfo{year}{1997}\natexlab{}.
\newblock \showarticletitle{Long short-term memory}.
\newblock \bibinfo{journal}{\emph{Neural computation}} \bibinfo{volume}{9},
  \bibinfo{number}{8} (\bibinfo{year}{1997}), \bibinfo{pages}{1735--1780}.
\newblock


\bibitem[\protect\citeauthoryear{Hu, Dong, Wang, and Sun}{Hu
  et~al\mbox{.}}{2020}]%
        {hu2020heterogeneous}
\bibfield{author}{\bibinfo{person}{Ziniu Hu}, \bibinfo{person}{Yuxiao Dong},
  \bibinfo{person}{Kuansan Wang}, {and} \bibinfo{person}{Yizhou Sun}.}
  \bibinfo{year}{2020}\natexlab{}.
\newblock \showarticletitle{Heterogeneous graph transformer}. In
  \bibinfo{booktitle}{\emph{Proceedings of The Web Conference 2020}}.
  \bibinfo{pages}{2704--2710}.
\newblock


\bibitem[\protect\citeauthoryear{Huang, Zhang, Rong, and Huang}{Huang
  et~al\mbox{.}}{2018}]%
        {huang2018adaptive}
\bibfield{author}{\bibinfo{person}{Wenbing Huang}, \bibinfo{person}{Tong
  Zhang}, \bibinfo{person}{Yu Rong}, {and} \bibinfo{person}{Junzhou Huang}.}
  \bibinfo{year}{2018}\natexlab{}.
\newblock \showarticletitle{Adaptive sampling towards fast graph representation
  learning}.
\newblock \bibinfo{journal}{\emph{Advances in neural information processing
  systems}}  \bibinfo{volume}{31} (\bibinfo{year}{2018}).
\newblock


\bibitem[\protect\citeauthoryear{Jin, Qin, Fang, Du, Zhang, Yu, Zhang, and
  Smola}{Jin et~al\mbox{.}}{2020}]%
        {jin2020efficient}
\bibfield{author}{\bibinfo{person}{Jiarui Jin}, \bibinfo{person}{Jiarui Qin},
  \bibinfo{person}{Yuchen Fang}, \bibinfo{person}{Kounianhua Du},
  \bibinfo{person}{Weinan Zhang}, \bibinfo{person}{Yong Yu},
  \bibinfo{person}{Zheng Zhang}, {and} \bibinfo{person}{Alexander~J Smola}.}
  \bibinfo{year}{2020}\natexlab{}.
\newblock \showarticletitle{An efficient neighborhood-based interaction model
  for recommendation on heterogeneous graph}. In
  \bibinfo{booktitle}{\emph{Proceedings of the 26th ACM SIGKDD International
  Conference on Knowledge Discovery \& Data Mining}}. \bibinfo{pages}{75--84}.
\newblock


\bibitem[\protect\citeauthoryear{Ke, Meng, Finley, Wang, Chen, Ma, Ye, and
  Liu}{Ke et~al\mbox{.}}{2017}]%
        {ke2017lightgbm}
\bibfield{author}{\bibinfo{person}{Guolin Ke}, \bibinfo{person}{Qi Meng},
  \bibinfo{person}{Thomas Finley}, \bibinfo{person}{Taifeng Wang},
  \bibinfo{person}{Wei Chen}, \bibinfo{person}{Weidong Ma},
  \bibinfo{person}{Qiwei Ye}, {and} \bibinfo{person}{Tie-Yan Liu}.}
  \bibinfo{year}{2017}\natexlab{}.
\newblock \showarticletitle{Lightgbm: A highly efficient gradient boosting
  decision tree}.
\newblock \bibinfo{journal}{\emph{Advances in neural information processing
  systems}}  \bibinfo{volume}{30} (\bibinfo{year}{2017}),
  \bibinfo{pages}{3146--3154}.
\newblock


\bibitem[\protect\citeauthoryear{Kipf and Welling}{Kipf and Welling}{2016}]%
        {kipf2016semi}
\bibfield{author}{\bibinfo{person}{Thomas~N Kipf} {and} \bibinfo{person}{Max
  Welling}.} \bibinfo{year}{2016}\natexlab{}.
\newblock \showarticletitle{Semi-supervised classification with graph
  convolutional networks}.
\newblock \bibinfo{journal}{\emph{arXiv preprint arXiv:1609.02907}}
  (\bibinfo{year}{2016}).
\newblock


\bibitem[\protect\citeauthoryear{Lian, Zhou, Zhang, Chen, Xie, and Sun}{Lian
  et~al\mbox{.}}{2018}]%
        {lian2018xdeepfm}
\bibfield{author}{\bibinfo{person}{Jianxun Lian}, \bibinfo{person}{Xiaohuan
  Zhou}, \bibinfo{person}{Fuzheng Zhang}, \bibinfo{person}{Zhongxia Chen},
  \bibinfo{person}{Xing Xie}, {and} \bibinfo{person}{Guangzhong Sun}.}
  \bibinfo{year}{2018}\natexlab{}.
\newblock \showarticletitle{xdeepfm: Combining explicit and implicit feature
  interactions for recommender systems}. In
  \bibinfo{booktitle}{\emph{Proceedings of the 24th ACM SIGKDD International
  Conference on Knowledge Discovery \& Data Mining}}.
  \bibinfo{pages}{1754--1763}.
\newblock


\bibitem[\protect\citeauthoryear{Lv, Jin, Yu, Sun, Lin, Yang, and Ng}{Lv
  et~al\mbox{.}}{2019}]%
        {lv2019sdm}
\bibfield{author}{\bibinfo{person}{Fuyu Lv}, \bibinfo{person}{Taiwei Jin},
  \bibinfo{person}{Changlong Yu}, \bibinfo{person}{Fei Sun},
  \bibinfo{person}{Quan Lin}, \bibinfo{person}{Keping Yang}, {and}
  \bibinfo{person}{Wilfred Ng}.} \bibinfo{year}{2019}\natexlab{}.
\newblock \showarticletitle{SDM: Sequential deep matching model for online
  large-scale recommender system}. In \bibinfo{booktitle}{\emph{Proceedings of
  the 28th ACM International Conference on Information and Knowledge
  Management}}. \bibinfo{pages}{2635--2643}.
\newblock


\bibitem[\protect\citeauthoryear{Lyu, Dong, Huo, and Ren}{Lyu
  et~al\mbox{.}}{2020}]%
        {lyu2020deep}
\bibfield{author}{\bibinfo{person}{Ze Lyu}, \bibinfo{person}{Yu Dong},
  \bibinfo{person}{Chengfu Huo}, {and} \bibinfo{person}{Weijun Ren}.}
  \bibinfo{year}{2020}\natexlab{}.
\newblock \showarticletitle{Deep Match to Rank Model for Personalized
  Click-Through Rate Prediction}. In \bibinfo{booktitle}{\emph{Proceedings of
  the AAAI Conference on Artificial Intelligence}}, Vol.~\bibinfo{volume}{34}.
  \bibinfo{pages}{156--163}.
\newblock


\bibitem[\protect\citeauthoryear{Ouyang, Zhang, Ren, Qi, Liu, and Du}{Ouyang
  et~al\mbox{.}}{2019}]%
        {ouyang2019representation}
\bibfield{author}{\bibinfo{person}{Wentao Ouyang}, \bibinfo{person}{Xiuwu
  Zhang}, \bibinfo{person}{Shukui Ren}, \bibinfo{person}{Chao Qi},
  \bibinfo{person}{Zhaojie Liu}, {and} \bibinfo{person}{Yanlong Du}.}
  \bibinfo{year}{2019}\natexlab{}.
\newblock \showarticletitle{Representation learning-assisted click-through rate
  prediction}.
\newblock \bibinfo{journal}{\emph{arXiv preprint arXiv:1906.04365}}
  (\bibinfo{year}{2019}).
\newblock


\bibitem[\protect\citeauthoryear{Qu, Bai, Zhang, Nie, and Tang}{Qu
  et~al\mbox{.}}{2019}]%
        {qu2019end}
\bibfield{author}{\bibinfo{person}{Yanru Qu}, \bibinfo{person}{Ting Bai},
  \bibinfo{person}{Weinan Zhang}, \bibinfo{person}{Jianyun Nie}, {and}
  \bibinfo{person}{Jian Tang}.} \bibinfo{year}{2019}\natexlab{}.
\newblock \showarticletitle{An end-to-end neighborhood-based interaction model
  for knowledge-enhanced recommendation}. In
  \bibinfo{booktitle}{\emph{Proceedings of the 1st International Workshop on
  Deep Learning Practice for High-Dimensional Sparse Data}}.
  \bibinfo{pages}{1--9}.
\newblock


\bibitem[\protect\citeauthoryear{Rendle}{Rendle}{2010}]%
        {rendle2010factorization}
\bibfield{author}{\bibinfo{person}{Steffen Rendle}.}
  \bibinfo{year}{2010}\natexlab{}.
\newblock \showarticletitle{Factorization machines}. In
  \bibinfo{booktitle}{\emph{2010 IEEE International conference on data
  mining}}. IEEE, \bibinfo{pages}{995--1000}.
\newblock


\bibitem[\protect\citeauthoryear{Rong, Bian, Xu, Xie, Wei, Huang, and
  Huang}{Rong et~al\mbox{.}}{2020a}]%
        {rong2020self}
\bibfield{author}{\bibinfo{person}{Yu Rong}, \bibinfo{person}{Yatao Bian},
  \bibinfo{person}{Tingyang Xu}, \bibinfo{person}{Weiyang Xie},
  \bibinfo{person}{Ying Wei}, \bibinfo{person}{Wenbing Huang}, {and}
  \bibinfo{person}{Junzhou Huang}.} \bibinfo{year}{2020}\natexlab{a}.
\newblock \showarticletitle{Self-supervised graph transformer on large-scale
  molecular data}.
\newblock \bibinfo{journal}{\emph{arXiv preprint arXiv:2007.02835}}
  (\bibinfo{year}{2020}).
\newblock


\bibitem[\protect\citeauthoryear{Rong, Huang, Xu, and Huang}{Rong
  et~al\mbox{.}}{2020b}]%
        {Rong2020DropEdge:}
\bibfield{author}{\bibinfo{person}{Yu Rong}, \bibinfo{person}{Wenbing Huang},
  \bibinfo{person}{Tingyang Xu}, {and} \bibinfo{person}{Junzhou Huang}.}
  \bibinfo{year}{2020}\natexlab{b}.
\newblock \showarticletitle{DropEdge: Towards Deep Graph Convolutional Networks
  on Node Classification}. In \bibinfo{booktitle}{\emph{International
  Conference on Learning Representations}}.
\newblock
\urldef\tempurl%
\url{https://openreview.net/forum?id=Hkx1qkrKPr}
\showURL{%
\tempurl}


\bibitem[\protect\citeauthoryear{Schlichtkrull, Kipf, Bloem, Van Den~Berg,
  Titov, and Welling}{Schlichtkrull et~al\mbox{.}}{2018}]%
        {schlichtkrull2018modeling}
\bibfield{author}{\bibinfo{person}{Michael Schlichtkrull},
  \bibinfo{person}{Thomas~N Kipf}, \bibinfo{person}{Peter Bloem},
  \bibinfo{person}{Rianne Van Den~Berg}, \bibinfo{person}{Ivan Titov}, {and}
  \bibinfo{person}{Max Welling}.} \bibinfo{year}{2018}\natexlab{}.
\newblock \showarticletitle{Modeling relational data with graph convolutional
  networks}. In \bibinfo{booktitle}{\emph{European semantic web conference}}.
  Springer, \bibinfo{pages}{593--607}.
\newblock


\bibitem[\protect\citeauthoryear{Seo, Kembhavi, Farhadi, and Hajishirzi}{Seo
  et~al\mbox{.}}{2016}]%
        {seo2016bidirectional}
\bibfield{author}{\bibinfo{person}{Minjoon Seo}, \bibinfo{person}{Aniruddha
  Kembhavi}, \bibinfo{person}{Ali Farhadi}, {and} \bibinfo{person}{Hannaneh
  Hajishirzi}.} \bibinfo{year}{2016}\natexlab{}.
\newblock \showarticletitle{Bidirectional attention flow for machine
  comprehension}.
\newblock \bibinfo{journal}{\emph{arXiv preprint arXiv:1611.01603}}
  (\bibinfo{year}{2016}).
\newblock


\bibitem[\protect\citeauthoryear{Vaswani, Shazeer, Parmar, Uszkoreit, Jones,
  Gomez, Kaiser, and Polosukhin}{Vaswani et~al\mbox{.}}{2017}]%
        {vaswani2017attention}
\bibfield{author}{\bibinfo{person}{Ashish Vaswani}, \bibinfo{person}{Noam
  Shazeer}, \bibinfo{person}{Niki Parmar}, \bibinfo{person}{Jakob Uszkoreit},
  \bibinfo{person}{Llion Jones}, \bibinfo{person}{Aidan~N Gomez},
  \bibinfo{person}{Lukasz Kaiser}, {and} \bibinfo{person}{Illia Polosukhin}.}
  \bibinfo{year}{2017}\natexlab{}.
\newblock \showarticletitle{Attention is all you need}.
\newblock \bibinfo{journal}{\emph{arXiv preprint arXiv:1706.03762}}
  (\bibinfo{year}{2017}).
\newblock


\bibitem[\protect\citeauthoryear{Velivckovic, Cucurull, Casanova, Romero, Lio,
  and Bengio}{Velivckovic et~al\mbox{.}}{2017}]%
        {velivckovic2017graph}
\bibfield{author}{\bibinfo{person}{Petar Velivckovic}, \bibinfo{person}{Guillem
  Cucurull}, \bibinfo{person}{Arantxa Casanova}, \bibinfo{person}{Adriana
  Romero}, \bibinfo{person}{Pietro Lio}, {and} \bibinfo{person}{Yoshua
  Bengio}.} \bibinfo{year}{2017}\natexlab{}.
\newblock \showarticletitle{Graph attention networks}.
\newblock \bibinfo{journal}{\emph{arXiv preprint arXiv:1710.10903}}
  (\bibinfo{year}{2017}).
\newblock


\bibitem[\protect\citeauthoryear{Wang, Fu, Fu, and Wang}{Wang
  et~al\mbox{.}}{2017}]%
        {wang2017deep}
\bibfield{author}{\bibinfo{person}{Ruoxi Wang}, \bibinfo{person}{Bin Fu},
  \bibinfo{person}{Gang Fu}, {and} \bibinfo{person}{Mingliang Wang}.}
  \bibinfo{year}{2017}\natexlab{}.
\newblock \showarticletitle{Deep \& cross network for ad click predictions}.
\newblock In \bibinfo{booktitle}{\emph{Proceedings of the ADKDD'17}}.
  \bibinfo{pages}{1--7}.
\newblock


\bibitem[\protect\citeauthoryear{Wang, He, Cao, Liu, and Chua}{Wang
  et~al\mbox{.}}{2019a}]%
        {wang2019kgat}
\bibfield{author}{\bibinfo{person}{Xiang Wang}, \bibinfo{person}{Xiangnan He},
  \bibinfo{person}{Yixin Cao}, \bibinfo{person}{Meng Liu}, {and}
  \bibinfo{person}{Tat-Seng Chua}.} \bibinfo{year}{2019}\natexlab{a}.
\newblock \showarticletitle{Kgat: Knowledge graph attention network for
  recommendation}. In \bibinfo{booktitle}{\emph{Proceedings of the 25th ACM
  SIGKDD International Conference on Knowledge Discovery \& Data Mining}}.
  \bibinfo{pages}{950--958}.
\newblock


\bibitem[\protect\citeauthoryear{Wang, Ji, Shi, Wang, Ye, Cui, and Yu}{Wang
  et~al\mbox{.}}{2019b}]%
        {wang2019heterogeneous}
\bibfield{author}{\bibinfo{person}{Xiao Wang}, \bibinfo{person}{Houye Ji},
  \bibinfo{person}{Chuan Shi}, \bibinfo{person}{Bai Wang},
  \bibinfo{person}{Yanfang Ye}, \bibinfo{person}{Peng Cui}, {and}
  \bibinfo{person}{Philip~S Yu}.} \bibinfo{year}{2019}\natexlab{b}.
\newblock \showarticletitle{Heterogeneous graph attention network}. In
  \bibinfo{booktitle}{\emph{The World Wide Web Conference}}.
  \bibinfo{pages}{2022--2032}.
\newblock


\bibitem[\protect\citeauthoryear{Wu, Zhang, Gao, He, Weng, Gao, and Chen}{Wu
  et~al\mbox{.}}{2019}]%
        {wu2019dual}
\bibfield{author}{\bibinfo{person}{Qitian Wu}, \bibinfo{person}{Hengrui Zhang},
  \bibinfo{person}{Xiaofeng Gao}, \bibinfo{person}{Peng He},
  \bibinfo{person}{Paul Weng}, \bibinfo{person}{Han Gao}, {and}
  \bibinfo{person}{Guihai Chen}.} \bibinfo{year}{2019}\natexlab{}.
\newblock \showarticletitle{Dual graph attention networks for deep latent
  representation of multifaceted social effects in recommender systems}. In
  \bibinfo{booktitle}{\emph{The World Wide Web Conference}}.
  \bibinfo{pages}{2091--2102}.
\newblock


\bibitem[\protect\citeauthoryear{Xiao, Ye, He, Zhang, Wu, and Chua}{Xiao
  et~al\mbox{.}}{2017}]%
        {xiao2017attentional}
\bibfield{author}{\bibinfo{person}{Jun Xiao}, \bibinfo{person}{Hao Ye},
  \bibinfo{person}{Xiangnan He}, \bibinfo{person}{Hanwang Zhang},
  \bibinfo{person}{Fei Wu}, {and} \bibinfo{person}{Tat-Seng Chua}.}
  \bibinfo{year}{2017}\natexlab{}.
\newblock \showarticletitle{Attentional factorization machines: Learning the
  weight of feature interactions via attention networks}.
\newblock \bibinfo{journal}{\emph{arXiv preprint arXiv:1708.04617}}
  (\bibinfo{year}{2017}).
\newblock


\bibitem[\protect\citeauthoryear{Ying, Cai, Luo, Zheng, Ke, He, Shen, and
  Liu}{Ying et~al\mbox{.}}{2021}]%
        {ying2021transformers}
\bibfield{author}{\bibinfo{person}{Chengxuan Ying}, \bibinfo{person}{Tianle
  Cai}, \bibinfo{person}{Shengjie Luo}, \bibinfo{person}{Shuxin Zheng},
  \bibinfo{person}{Guolin Ke}, \bibinfo{person}{Di He},
  \bibinfo{person}{Yanming Shen}, {and} \bibinfo{person}{Tie-Yan Liu}.}
  \bibinfo{year}{2021}\natexlab{}.
\newblock \showarticletitle{Do Transformers Really Perform Bad for Graph
  Representation?}
\newblock \bibinfo{journal}{\emph{arXiv preprint arXiv:2106.05234}}
  (\bibinfo{year}{2021}).
\newblock


\bibitem[\protect\citeauthoryear{Zhang, Song, Huang, Swami, and Chawla}{Zhang
  et~al\mbox{.}}{2019}]%
        {zhang2019heterogeneous}
\bibfield{author}{\bibinfo{person}{Chuxu Zhang}, \bibinfo{person}{Dongjin
  Song}, \bibinfo{person}{Chao Huang}, \bibinfo{person}{Ananthram Swami}, {and}
  \bibinfo{person}{Nitesh~V Chawla}.} \bibinfo{year}{2019}\natexlab{}.
\newblock \showarticletitle{Heterogeneous graph neural network}. In
  \bibinfo{booktitle}{\emph{Proceedings of the 25th ACM SIGKDD International
  Conference on Knowledge Discovery \& Data Mining}}.
  \bibinfo{pages}{793--803}.
\newblock


\bibitem[\protect\citeauthoryear{Zhang, Zhang, Xia, and Sun}{Zhang
  et~al\mbox{.}}{2020}]%
        {zhang2020graph}
\bibfield{author}{\bibinfo{person}{Jiawei Zhang}, \bibinfo{person}{Haopeng
  Zhang}, \bibinfo{person}{Congying Xia}, {and} \bibinfo{person}{Li Sun}.}
  \bibinfo{year}{2020}\natexlab{}.
\newblock \showarticletitle{Graph-bert: Only attention is needed for learning
  graph representations}.
\newblock \bibinfo{journal}{\emph{arXiv preprint arXiv:2001.05140}}
  (\bibinfo{year}{2020}).
\newblock


\bibitem[\protect\citeauthoryear{Zhou, Mou, Fan, Pi, Bian, Zhou, Zhu, and
  Gai}{Zhou et~al\mbox{.}}{2019}]%
        {zhou2019deep}
\bibfield{author}{\bibinfo{person}{Guorui Zhou}, \bibinfo{person}{Na Mou},
  \bibinfo{person}{Ying Fan}, \bibinfo{person}{Qi Pi}, \bibinfo{person}{Weijie
  Bian}, \bibinfo{person}{Chang Zhou}, \bibinfo{person}{Xiaoqiang Zhu}, {and}
  \bibinfo{person}{Kun Gai}.} \bibinfo{year}{2019}\natexlab{}.
\newblock \showarticletitle{Deep interest evolution network for click-through
  rate prediction}. In \bibinfo{booktitle}{\emph{Proceedings of the AAAI
  conference on artificial intelligence}}, Vol.~\bibinfo{volume}{33}.
  \bibinfo{pages}{5941--5948}.
\newblock


\bibitem[\protect\citeauthoryear{Zhou, Zhu, Song, Fan, Zhu, Ma, Yan, Jin, Li,
  and Gai}{Zhou et~al\mbox{.}}{2018}]%
        {zhou2018deep}
\bibfield{author}{\bibinfo{person}{Guorui Zhou}, \bibinfo{person}{Xiaoqiang
  Zhu}, \bibinfo{person}{Chenru Song}, \bibinfo{person}{Ying Fan},
  \bibinfo{person}{Han Zhu}, \bibinfo{person}{Xiao Ma},
  \bibinfo{person}{Yanghui Yan}, \bibinfo{person}{Junqi Jin},
  \bibinfo{person}{Han Li}, {and} \bibinfo{person}{Kun Gai}.}
  \bibinfo{year}{2018}\natexlab{}.
\newblock \showarticletitle{Deep interest network for click-through rate
  prediction}. In \bibinfo{booktitle}{\emph{Proceedings of the 24th ACM SIGKDD
  International Conference on Knowledge Discovery \& Data Mining}}.
  \bibinfo{pages}{1059--1068}.
\newblock


\end{thebibliography}
\newpage
\appendix

\eat{
\section{Appendix}
\subsection{The notations}
Table~\ref{tab:notation} depicts the frequently-used notations we used in the paper.
\begin{table}[h!]
\vspace{-1ex}
\caption{The Notations}
\vspace{-2ex}
 \resizebox{0.40\textwidth}{!}{
\begin{tabular}{l|l}
\hline
Notation                                            & Description                           \\ \hline
$\mathcal{G}(\mathcal{N}, \mathcal{E},\mathcal{T}_\mathcal{V},\mathcal{T}_\mathcal{E})$ & The Heterogeneous Information Network \\
$\mathcal{N}$& Node set\\
$\mathcal{E}$& Edge set\\
$\mathcal{T_N}$& Node type set \\
$\mathcal{T_E}$& Node type set \\
$\mathcal{F}$& Feature group set\\
$u$ &  Target user   \\
$v$ &Target item \\      
$\mathcal{N}_{uv}$& Neighbouring nodes of $u$ and $v$ \\
$y_{uv}$& The label of the $u$-$v$ pair \\
$\mathcal{F}_{uv}$& Set of node feature vectors in $\mathcal{N}_{uv}$ \\
$\BC$& Context feature set of $u$-$v$ \\
$t(i)$ & The node type of  node $i$ \\
$\f$ & The original feature vector of a node \\
$\x$ & The embedded feature vector of a node \\
$\h_i$& Embedding of node $i$ before a MSA layer\\
$\h_i'$& Embedding of node $i$ after a MSA layer\\
$\g_{uv}$& Neighbourhood embedding of $u$-$v$\\
$\BQ,\BK,\BV,\BW^O$ & Parameter matrices \\
$f_m(\cdot)$ & The masking function \\
\hline
\end{tabular}}
\label{tab:notation}
\end{table}
}

\end{document}